\documentclass{aastex701}

\makeatletter
\newcommand\raa{\ref@jnl{RAA}}
\makeatother

\usepackage{multirow}
\usepackage{booktabs}

\shorttitle{HXI-DLA2: Physics-Constrained Deep Learning Network for the ASO-S Hard X-ray Imager}
\shortauthors{Zou et al.}

\graphicspath{{./}{img/}}

\begin{document}

\title{HXI-DLA2: A Physics-Constrained Deep Learning Algorithm for the ASO-S Hard X-ray Imager}

%% Author block: each author requires \author, \affiliation, and \email.
%% The optional argument provides ORCID, surname, and given name metadata.

\author[orcid=0000-0001-7607-2594,gname=SiZhong,sname=Zou]{Sizhong Zou}
\affiliation{Yunnan Observatories, Chinese Academy of Sciences, Kunming 650216, China}
\affiliation{University of Chinese Academy of Sciences, Beijing 100049, China}
\email{zousizhong@ynao.ac.cn}

\author[orcid=0000-0003-2714-6811,gname=Hui,sname=Liu]{Hui Liu}
\affiliation{Yunnan Observatories, Chinese Academy of Sciences, Kunming 650216, China}
\affiliation{Yunnan Key Laboratory of Solar Physics and Space Science, Kunming 650216, China}
\email{liuhui@ynao.ac.cn}

\author[orcid=0000-0002-4241-9921,gname=Yang,sname=Su]{Yang Su}
\affiliation{Key Laboratory of Dark Matter and Space Astronomy, Purple Mountain Observatory, Chinese Academy of Sciences, Nanjing 210034, China}
\email{yangsu@pmo.ac.cn}

\author[orcid=0000-0002-3804-7395,gname=JunChao,sname=Hong]{Junchao Hong}
\affiliation{Yunnan Observatories, Chinese Academy of Sciences, Kunming 650216, China}
\affiliation{Yunnan Key Laboratory of Solar Physics and Space Science, Kunming 650216, China}
\email{hjcsolar@ynao.ac.cn}

\author[orcid=0000-0002-8300-3199,gname=Bin,sname=Wang]{Bin Wang}
\affiliation{Yunnan Observatories, Chinese Academy of Sciences, Kunming 650216, China}
\affiliation{Yunnan Key Laboratory of Solar Physics and Space Science, Kunming 650216, China}
\email{wangbin@ynao.ac.cn}

\author[orcid=0000-0001-5279-3266,gname=Wei,sname=Chen]{Wei Chen}
\affiliation{Key Laboratory of Dark Matter and Space Astronomy, Purple Mountain Observatory, Chinese Academy of Sciences, Nanjing 210034, China}
\email{chenwei@pmo.ac.cn}

\author[orcid=0000-0001-8950-3875,gname=KaiFan,sname=Ji]{Kaifan Ji}
\affiliation{Yunnan Observatories, Chinese Academy of Sciences, Kunming 650216, China}
\affiliation{Yunnan Key Laboratory of Solar Physics and Space Science, Kunming 650216, China}
\email{jkf@ynao.ac.cn}

\author[orcid=0000-0001-7575-5449,gname=ZhenYu,sname=Jin]{Zhenyu Jin}
\affiliation{Yunnan Observatories, Chinese Academy of Sciences, Kunming 650216, China}
\affiliation{Yunnan Key Laboratory of Solar Physics and Space Science, Kunming 650216, China}
\email{kim@ynao.ac.cn}

%% AAS Journals have a 250 word limit for the abstract.
\begin{abstract}

Solar flare hard X-ray imaging is a key diagnostic of flare energy release and electron acceleration. The Hard X-ray Imager (HXI) aboard ASO-S compresses the two-dimensional source distribution into counts measured by 91 sub-collimators, making image reconstruction an inherently underdetermined inverse problem. Conventional algorithms such as CLEAN rely on point-source priors and manual tuning, whereas recent deep-learning methods offer no guarantee that their reconstructions obey the instrument's modulation-sampling forward equation. In this work we show that the counts decompose into two nearly decoupled quantities---the counts average energy, which tracks the total source flux, and the normalized counts distribution, which encodes the source spatial structure---and we exploit this property to construct a physics-constrained network, the Hard X-ray Imager Deep Learning Algorithm 2 (HXI-DLA2). Non-negativity and exact counts-average-energy closure are enforced at the network output, while a distribution-consistency loss aligns the re-projected counts with the measurement, so that the reconstruction satisfies the forward equation by construction. Tests on simulated Gaussian sources, observed soft X-ray morphologies, and a real HXI flare event show two main improvements over existing methods: the limiting resolvable dynamic range of double sources is pushed well beyond that of conventional imaging algorithms and our previous method; and complex morphologies on which prior reconstructions degrade, such as ring-like and diffuse structures, are reliably reconstructed, with the real-event result consistent with contemporaneous SDO/AIA imaging. Embedding the instrumental forward equation as a hard constraint while learning source priors from data offers a general inversion framework for modulation imaging.

\end{abstract}

%% The AAS Journals use Unified Astronomy Thesaurus (UAT) concepts.


\section{Introduction}

Solar flares are among the most violent energy-release phenomena in the solar system; their energy release involves magnetic reconnection, particle acceleration, and plasma heating. In the standard flare model, coronal magnetic reconnection converts magnetic energy into kinetic energy, accelerating nonthermal electrons that propagate along magnetic field lines into the lower atmosphere, where they interact with the dense plasma and produce bremsstrahlung emission that forms the characteristic radiation in the hard X-ray (HXR, usually photons above 10 keV) band. Hard X-ray imaging is therefore a key diagnostic of the location of flare energy release, the electron acceleration mechanism, and the particle transport paths.

Since the 1990s, the development of space-borne hard X-ray imagers has greatly advanced solar high-energy physics. Yohkoh/HXT achieved the first systematic solar hard X-ray imaging observations \citep{Kosugi1991}; RHESSI provided a large number of high-cadence flare hard X-ray images through rotating modulation collimator imaging \citep{Hurford2002}; and Solar Orbiter/STIX further extended the vantage points of solar high-energy imaging \citep{Krucker2020, Muller2020}. These instruments demonstrate that hard X-ray imaging not only reveals the spatial distribution of flare energy sources but also enables imaging spectroscopy to infer electron spectra and energy deposition.

China's first dedicated solar science mission, the Advanced Space-based Solar Observatory (ASO-S), was successfully launched on 2022 October 9 \citep{Gan2022}, with the core science objective of ``one magnetic field plus two eruptions''---simultaneous observations of the solar magnetic field, solar flares, and coronal mass ejections. ASO-S carries three main payloads: the Full-disk vector MagnetoGraph (FMG), the Lyman-alpha Solar Telescope (LST), and the Hard X-ray Imager (HXI). HXI is dedicated to observing the spectra and images of solar flare hard X-ray emission and is crucial for studying the acceleration of energetic electrons. HXI adopts a Fourier-transform imaging technique based on bi-grid modulation \citep{Zhang2023}, consisting of 91 tungsten-grid sub-collimators and LaBr$_3$ scintillator detectors; each sub-collimator contains a front and a rear grid, and different pitches and position angles sample different spatial frequencies. The 91 detectors cover spatial scales from 3\arcsec\ to 105\arcsec\ (see Table~\ref{tab:hxi_detector_summary}), and inversion algorithms ultimately reconstruct solar flare hard X-ray images with a spatial resolution of about 3\arcsec.

\begin{table*}[ht!]
\centering
\small
\setlength{\tabcolsep}{5pt}
\caption{Grouped configuration and physical parameters of the 91 HXI detectors.}
\label{tab:hxi_detector_summary}
\begin{tabular}{lcccccccccc}
\toprule
Pitch ($\mu$m) & 36 & 52 & 76 & 108 & 156 & 224 & 344 & 524 & 800 & 1224 \\
Angular resolution ($''$) & 3.1 & 4.5 & 6.5 & 9.3 & 13.4 & 19.3 & 29.6 & 45.1 & 68.8 & 105.2 \\
\midrule
Detectors at phase $\varphi=0^\circ$ & 4 & 5 & 5 & 5 & 5 & 5 & 5 & 5 & 3 & 3 \\
Detectors at phase $\varphi=90^\circ$ & 4 & 5 & 5 & 5 & 5 & 5 & 5 & 5 & 3 & 2 \\
Detectors at phase $\varphi=120^\circ$ & 0 & 0 & 0 & 0 & 0 & 0 & 0 & 0 & 0 & 1 \\
Detectors at phase $\varphi=240^\circ$ & 0 & 0 & 0 & 0 & 0 & 0 & 0 & 0 & 0 & 1 \\
\midrule
\multirow{5}{*}{Position angles ($^\circ$)} & 25 & 5 & 32 & 23 & 14 & 5 & 23 & 5 & 23 & 53 \\
 & 70 & 41 & 68 & 59 & 50 & 41 & 59 & 41 & 83 & 113 \\
 & 115 & 77 & 104 & 95 & 86 & 77 & 95 & 77 & 143 & 173 \\
 & 160 & 113 & 140 & 131 & 122 & 113 & 131 & 113 & & \\
 &     & 149 & 176 & 167 & 158 & 149 & 167 & 149 & & \\
\bottomrule
\end{tabular}
\end{table*}

Mathematically, HXI imaging is a linear projection: each of the 91 counts is a weighted sum of the source image over the corresponding modulation pattern (Equation~(\ref{eq:forward}) in Section 2), so the two-dimensional source is compressed into a 91-dimensional counts vector and imaging becomes an underdetermined inverse problem of \textbf{recovering a high-dimensional spatial distribution from low-dimensional measurements}. Position, scale, and intensity are entangled in the counts, the solution is not unique, and physically credible source images can be recovered only with suitable priors and inversion algorithms. Indeed, indirect modulation imaging can measure only a limited number of spatial-frequency components, and this sampling limitation fundamentally constrains the morphological complexity of the source that can be reconstructed \citep{Hurford2002, Krucker2020}: for extended or multi-component complex sources, the available information is insufficient to uniquely determine the spatial distribution---this is precisely the common challenge faced by modulation-imaging instruments.

HXI inversion algorithms fall into two categories: direct inversion based on the modulation patterns and visibility-based imaging. In direct inversion, back-projection multiplies the counts by the modulation patterns and accumulates them to obtain a ``dirty image''; because detector phase relations and noise properties are not accounted for, strong negative sidelobes and modulation fringes remain outside the source region, so the dirty image cannot be used directly for science. The CLEAN algorithm originates from radio interferometric imaging \citep{Hogbom1974,Clark1980}; under a point-source assumption, it iteratively locates the maximum residual, subtracts a clean beam, and truncates negative values to zero to suppress sidelobes. This assumption is effective for isolated point sources, but for flares with continuous loop-like or extended complex morphologies the reconstruction is often fragmented, unsmooth, and grainy. Maximum-entropy methods (MEM\_GE, MEM\_NJIT) \citep{Cornwell1985} and expectation--maximization (EM) improve resolution through different priors and have become standard tools for HXI data processing. Visibility-based imaging methods first pair the 91 real counts into 45 complex visibilities and then reconstruct the source image via Fourier inversion; this approach is computationally efficient, but the visibility synthesis process loses part of the information among the counts and is sensitive to noise. Overall, conventional algorithms rely on strong priors and manual parameter tuning, have limited capability for morphologically complex flares (multiple sources, loop-top sources, extended flares), and cannot exploit the statistical regularities in historical observations. New inversion approaches that combine data-driven models with physical constraints have therefore become an important direction for HXI imaging research.

In recent years, deep learning has shown strong potential for solar physics data processing. For HXI imaging, an earlier work by our group \citep{Xia2024} proposed the HXI-DLA neural network imaging algorithm, the first systematic application of deep learning to solar high-energy imaging inversion. HXI-DLA takes the 91-dimensional counts as input and the source-image pixel intensities as output, extracting spatial features with MLPs and convolutional networks; its reconstructions are reliable for compact sources, but the performance degrades for extended sources. \citet{Selcuk2025} proposed a Fourier Convolutional Decoder (FCD) for STIX, taking Fourier components directly as input to achieve millisecond-level inference with far fewer parameters than HXI-DLA. These two works represent the main direction of deep-learning-based modulation imaging: learning a mapping from counts to image in a data-driven manner. However, whether through the pixel-level image supervision of HXI-DLA or the Fourier-component regression of FCD, neither approach imposes an explicit constraint from the forward physical model---there is no mathematical guarantee that the reconstruction strictly satisfies the optical sampling forward equation. On the other hand, physics-informed neural networks \citep[PINNs;][]{Raissi2019} have shown effectiveness in astronomical image recovery: \citet{Ni2025} applied the PINN framework to astronomical image deconvolution, demonstrating the benefit of physical constraints; however, the constraint form is designed for deconvolution and is not applicable to the bi-grid modulation forward model of HXI. \citet{DiazBaso2025} applied PINNs to solar magnetic field spectropolarimetric inversion (Milne--Eddington approximation), embedding the instrumental PSF as a physical constraint in the network and achieving noise-robust inversion over an extended field of view; however, their constraint targets spectral-line profile fitting rather than modulation-imaging counts consistency.

The central question of this paper is: among the 91-dimensional counts of modulation imaging, which physical quantities can be uniquely determined and which depend on the source spatial distribution, and how can this structure be translated into executable inversion constraints in a deep network. Inspired by PINNs, this work builds upon our group's earlier HXI-DLA work and continues the development of image reconstruction methods for the HXI instrument. Starting from the modulation principles of the optical system, it embeds the HXI optical sampling forward equation as a mathematically known prior into the network architecture and optimization process, and designs an imaging network HXI-DLA2 that has both the nonlinear fitting power of deep networks and the constraints of the forward equation (non-negativity and energy consistency). The main contributions are:

\begin{enumerate}
\item \textbf{Counts decoupling in the measurement domain} (Section~2): single-factor analysis of the modulation forward model shows that the counts average energy scales with the total source flux, while the normalized counts distribution encodes the source spatial structure; the two are approximately separable, with the residual coupling a higher-order effect. This decoupling provides two independently enforceable constraints for the underdetermined inversion.

\item \textbf{Physics-constrained inversion network} (Section~3): HXI-DLA2 embeds the decoupled quantities as complementary constraints---ReLU non-negativity and enforced counts-average-energy rescaling at the output, and a counts-distribution-consistency loss during training---so that the reconstruction obeys the instrument's forward equation by construction.

\item \textbf{Multi-level validation} (Sections~3--5): experiments on Gaussian simulated sources, soft X-ray complex morphologies, and a real HXI flare event cover single, double, and ring sources; double sources remain resolved at peak ratios up to the extreme 1:50---a regime unresolved by our previous method---and the real-event reconstruction is cross-checked against contemporaneous SDO/AIA observations.

\item \textbf{Method positioning} (Sections~3 and 5): the forward equation enters the reconstruction as an explicit constraint rather than implicit supervision, removing the dependence on strong morphological priors and manual parameter tuning; the framework is in principle applicable to other modulation-imaging instruments.
\end{enumerate}

The paper is organized as follows. Section 2 analyzes the HXI modulation imaging physics and establishes the counts decoupling relationship; Section 3 presents the HXI-DLA2 network architecture, the physics-inspired loss function, the training strategy, and the Gaussian simulated source experiments; Section 4 reports the reconstruction results on soft X-ray complex morphologies and a real HXI observation event; Section 5 extends the double-source experiments to the extreme dynamic range, discusses the physical nature and capability limits of the reconstruction, and concludes with a summary and outlook.

\section{HXI Modulation Imaging and Counts Decoupling}

Starting from the physical model of HXI modulation imaging, this section derives the mathematical form of the forward equation, reveals the independent effects of position, scale, and intensity on the counts through single-factor analysis, and finally proposes a counts average-energy--distribution decoupling strategy, providing the theoretical basis for the network design in Section 3.

ASO-S/HXI performs Fourier-transform imaging with bi-grid modulation collimators \citep{Zhang2023, Hurford2002}. The $i$-th sub-collimator consists of front and rear tungsten grids whose transmission functions can be approximated in cosine form:
\begin{equation}
T_{i}^{(1)}(x,y) = \frac{1}{2}\Bigl[1 + \cos\bigl(2\pi \mathbf{k}_i \cdot \mathbf{r} + \phi_i^{(1)}\bigr)\Bigr], \quad
T_{i}^{(2)}(x,y) = \frac{1}{2}\Bigl[1 + \cos\bigl(2\pi \mathbf{k}_i \cdot \mathbf{r} + \phi_i^{(2)}\bigr)\Bigr],
\label{eq:transmission}
\end{equation}
where $\mathbf{r}=(x,y)$, the spatial frequency vector is $\mathbf{k}_i = (\cos\theta_i/p_i, \sin\theta_i/p_i)$, $p_i$ is the grid pitch, $\theta_i$ the position angle, and $\phi_i^{(1,2)}$ the phases of the front and rear grids. The overall transmission of the sub-collimator is $P_i(x,y) = T_{i}^{(1)}(x,y) \cdot T_{i}^{(2)}(x,y)$. For an incident solar hard X-ray source image $I(x,y)\ge 0$, the counts recorded by the $i$-th detector are
\begin{equation}
C_i = \sum_{x,y} P_i(x,y)\, I(x,y), \quad i=1,\ldots,91.
\label{eq:forward}
\end{equation}
Equation~(\ref{eq:forward}) constitutes the linear forward model of HXI imaging, compressing the two-dimensional source image into 91 scalar counts. The model can be decomposed into DC and AC components: writing the spatial average of the transmission function as $\bar{P}_i = \langle P_i \rangle_{x,y}$, Equation~(\ref{eq:forward}) becomes
\begin{equation}
C_i = \bar{P}_i \sum_{x,y} I(x,y) + \sum_{x,y} \bigl[P_i(x,y) - \bar{P}_i\bigr] I(x,y) = \bar{P}_i \cdot E + \text{AC}_i,
\label{eq:dcac}
\end{equation}
where $E = \sum_{x,y} I(x,y)$ is the total source energy (equivalently, the total source flux, i.e., the integrated image intensity). The first term is the \textbf{DC component}, proportional to the total source energy and reflecting the common response of the 91 detectors to the total incident X-ray flux; $\bar{P}_i$ is nearly constant across channels (all sub-collimators have approximately the same average transmission), so the DC component does not vary with the channel index $i$ and is determined solely by the total source energy. The second term is the \textbf{AC component}, produced by the spatial cosine modulation of the transmission functions; it varies from channel to channel because the modulation pattern $P_i(x,y) - \bar{P}_i$ differs among detectors, and it depends on the overlap of the source distribution $I(x,y)$ with the nodes and antinodes of $P_i(x,y)$, thus encoding the response at different spatial frequencies.

Because the pitches $p_i$ and position angles $\theta_i$ differ among detectors, the spatial frequencies $\|\mathbf{k}_i\| = 1/p_i$ differ among channels: smaller pitches correspond to higher spatial frequencies, and the AC component is sensitive to fine structure, whereas larger pitches mainly probe large-scale distributions.

To isolate the effect of each source factor on the counts, we perform a single-factor analysis with Gaussian simulated sources. Gaussian simulated sources have simple analytic forms, controllable parameters (position, standard deviation, peak intensity), and physical similarity to real compact flare sources, making them the standard benchmark for validating imaging algorithms \citep{Xia2024}. Given the source parameters, the corresponding 91-dimensional counts can be computed exactly via Equation~(\ref{eq:forward}), making Gaussian sources ideal tools for validating physical laws and decoupling strategies. Fixing two of the three key factors and varying the remaining one, we derive the independent effects of position, areal scale, and intensity on the counts.

\begin{figure*}[ht!]
\centering
\textbf{(a)}\\
\includegraphics[width=0.72\textwidth]{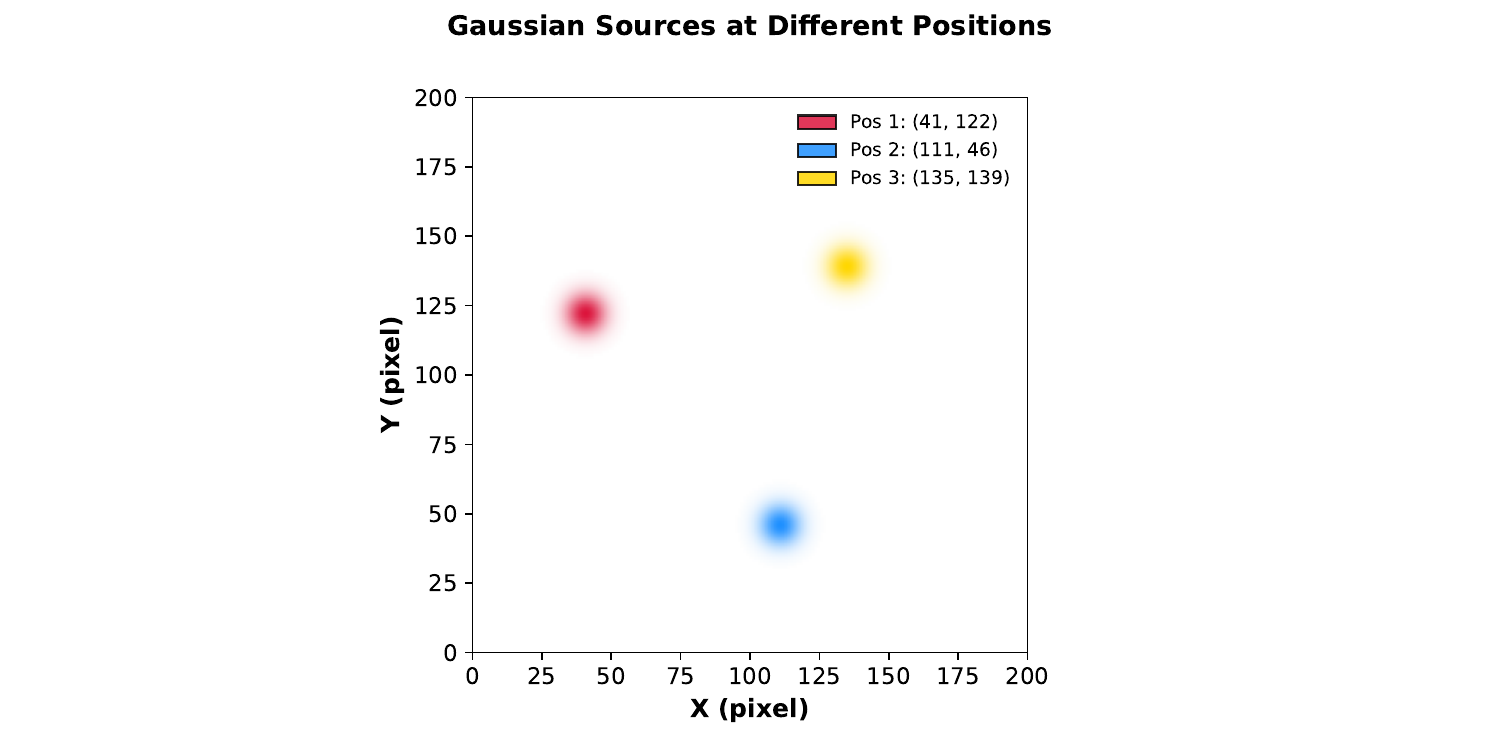}
\\[2pt]
\textbf{(b)}\\
\includegraphics[width=0.72\textwidth]{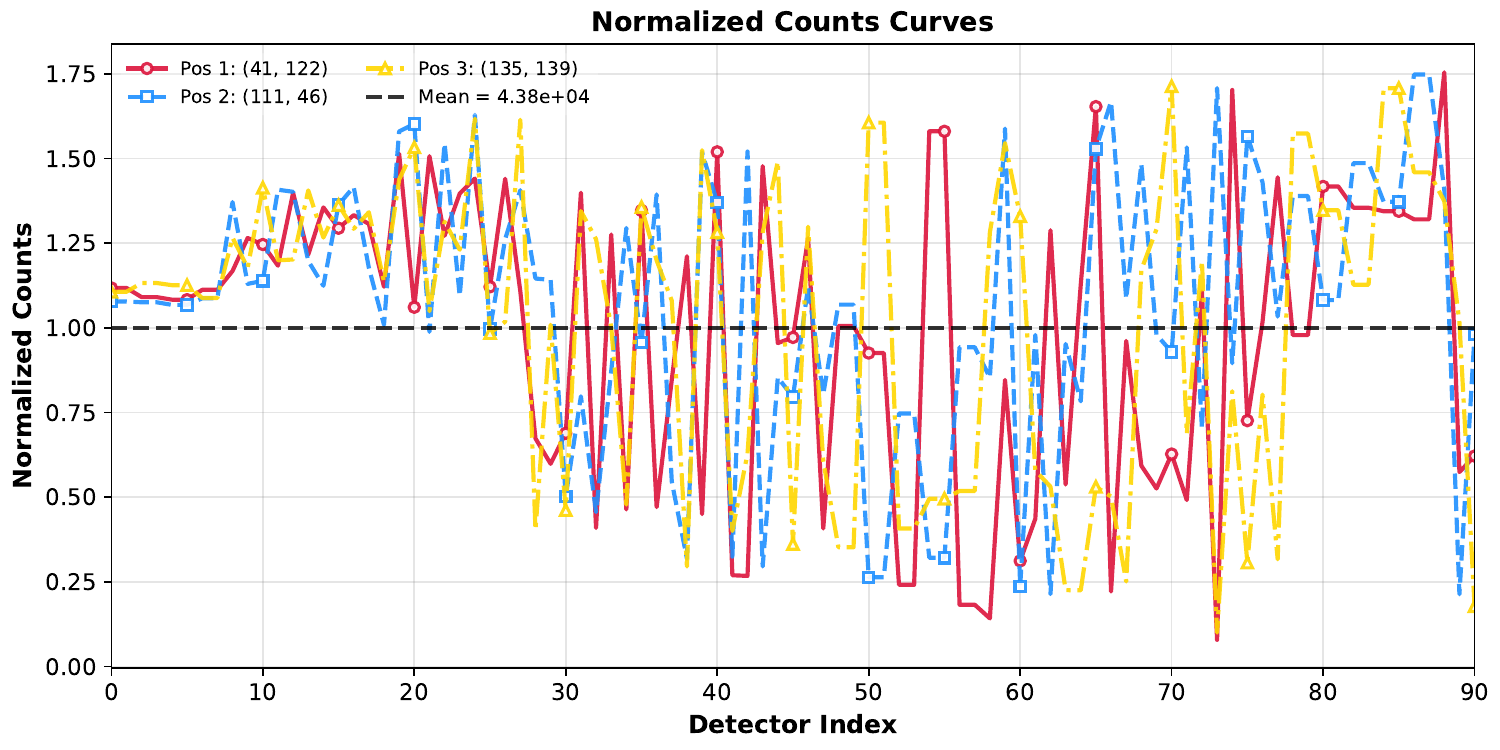}
\caption{(a) Source-position analysis: three Gaussian simulated sources with identical scale and peak intensity but different positions. (b) The corresponding 91-dimensional counts responses: the three sources share the same counts average energy but markedly different counts distributions, confirming that position only changes the phase distribution of the counts without affecting the total source energy.}
\label{fig:2_pos_intensity}
\end{figure*}

\textbf{Position: counts distribution and phase information.} Fixing the intensity and scale, only the spatial position is varied, as shown in Figure~\ref{fig:2_pos_intensity}(a). The counts average energy at different positions are equal (constant total source energy), but the normalized counts distributions change significantly. Physically, when the point-source position $\mathbf{r}_0$ moves, the response of the $i$-th detector is redistributed according to the phase $\mathbf{k}_i\cdot\mathbf{r}_0$, changing the relative weights among channels. The counts distribution (i.e., the phase distribution) thus carries the positional information of the source, as shown in Figure~\ref{fig:2_pos_intensity}(b).

\begin{figure*}[ht!]
\centering
\includegraphics[width=0.85\textwidth]{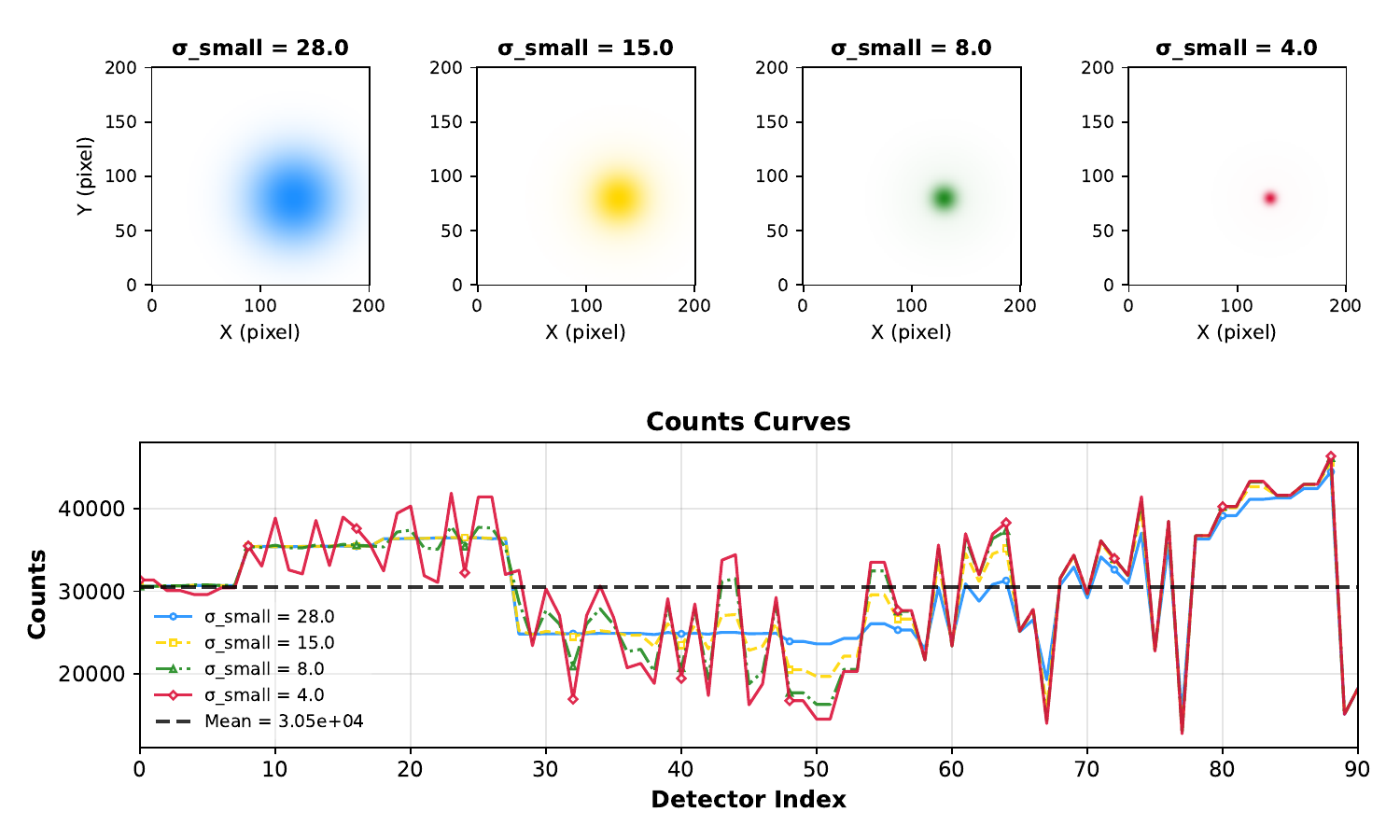}
\caption{Source-scale analysis. Four Gaussian simulated sources with identical total source energy but different scales, and the corresponding 91-dimensional counts responses. The counts average energies are nearly identical, the low-frequency responses are similar in shape, while the high-frequency channels differ significantly, confirming that scale only affects the high-frequency components of the counts and is decoupled from the total source energy.}
\label{fig:2_scale}
\end{figure*}

\textbf{Areal scale: mid- and high-frequency components of the counts.} Fixing the intensity and position, the Gaussian standard deviation $\sigma$ is varied. The counts average energy changes little, but the responses of small-pitch (high-frequency) detectors change markedly: a large source spans several high-frequency modulation periods, covering both crests and troughs so that the high-frequency modulation cancels in the spatial integral and the response weakens; a small source covers only a local region matched to the fine pitch of high-frequency detectors and excites significant modulation contrast. The mid- and high-frequency components of the counts are therefore sensitive to the source scale, while the average energy is decoupled from the scale, as shown in Figure~\ref{fig:2_scale}.

\textbf{Intensity: counts average energy and total source energy.} Fixing the position and standard deviation $\sigma$ of the Gaussian source (i.e., keeping the source area unchanged), only the peak intensity (i.e., the total source energy $E=\sum_{x,y}I(x,y)$) is varied. By the linearity of the projection integral in Equation~(\ref{eq:forward}), an overall scaling $I\to\alpha I$ scales the counts proportionally, $C_i\to\alpha C_i$. The counts average energy $\bar{C}=\frac{1}{91}\sum_{i=1}^{91}C_i$ is therefore proportional to the total source energy, while the normalized counts distribution $\hat{C}_i=C_i/\bar{C}$ remains unchanged. It should be noted that this proportionality holds under the premise of fixed source area; when the source scale changes, the non-vanishing residual of the AC component in Equation~(\ref{eq:dcac}) causes a slight deviation in the counts average energy (see Figure~\ref{fig:2_scale}). This analysis shows that, under the condition of fixed source area, the counts average energy is a direct measure of the total source energy and can serve as an independent physical constraint.

Summarizing the single-factor analyses, the counts average energy and distribution can be approximately decoupled: the former is determined by the total source energy, the latter by the spatial distribution. It should be emphasized that this decoupling is an ideal conclusion established under single-factor control experiments; in real flare sources, intensity, position, and scale often vary simultaneously, leading to weak coupling between the two. Nevertheless, this weak coupling is a higher-order effect relative to the dominant trends, and the average-energy--distribution decoupling can still serve as a first-order approximation providing two separately enforceable physical constraints for the underdetermined inversion.

The single-source analysis above extends naturally to multi-source cases. Real flare sources often consist of multiple superposed sub-sources. By the linearity of the forward model (\ref{eq:forward}), the total counts equal the algebraic sum of the sub-source counts: the counts average energy remains additive under superposition, while the counts distribution is modulated by the differing spatial distributions. The average-energy--distribution decoupling therefore applies to single and multiple sources alike.

Based on the above analysis, we decompose the 91-dimensional counts vector into two complementary physical quantities:
\begin{equation}
\bar{C} = \frac{1}{91}\sum_{i=1}^{91} C_i, \qquad \hat{\mathbf{C}} = \frac{\mathbf{C}}{\bar{C}}.
\label{eq:decouple}
\end{equation}
Here $\bar{C}$ is the \textbf{counts average energy}, which responds linearly to the source intensity, corresponds to the DC component, and serves as the scalar constraint on the total source energy; $\hat{\mathbf{C}}$ is the \textbf{counts distribution}, sensitive to position, scale, and morphology but independent of the total intensity, corresponding to the AC component and serving as the vector constraint on the spatial distribution.

This decoupling strategy provides the theoretical basis for the network design in Section 3: the counts average energy $\bar{C}$ is mapped to an energy constraint at the network output, and the counts distribution $\hat{\mathbf{C}}$ is mapped to a spatial-distribution constraint in the loss function.

\section{HXI-DLA2 Network Design and Validation}

Based on the counts average-energy--distribution decoupling established in Section~2 (Equation~\ref{eq:decouple}), this section constructs the HXI-DLA2 network. Inspired by the PINN idea of embedding physical equations into the loss function \citep{Raissi2019}, this work takes the forward equation (Equation~\ref{eq:forward}) as the governing equation and embeds its physical constraints into the network architecture and loss function, comprising three layers: a \textbf{non-negativity constraint} $I(x,y)\ge 0$, an \textbf{energy-consistency constraint} (the re-projected counts average energy must strictly equal the measured $\bar{C}$), and a \textbf{distribution-consistency constraint} (the re-projected normalized counts distribution is driven toward the measured $\hat{\mathbf{C}}$ through the loss function). The network architecture (Section~3.1), the loss function (Section~3.2), and the training strategy (Section~3.3) are described in turn.

\subsection{Network Architecture and Physics-Constraint Embedding}

\begin{figure*}[ht!]
\centering
\includegraphics[width=0.88\textwidth]{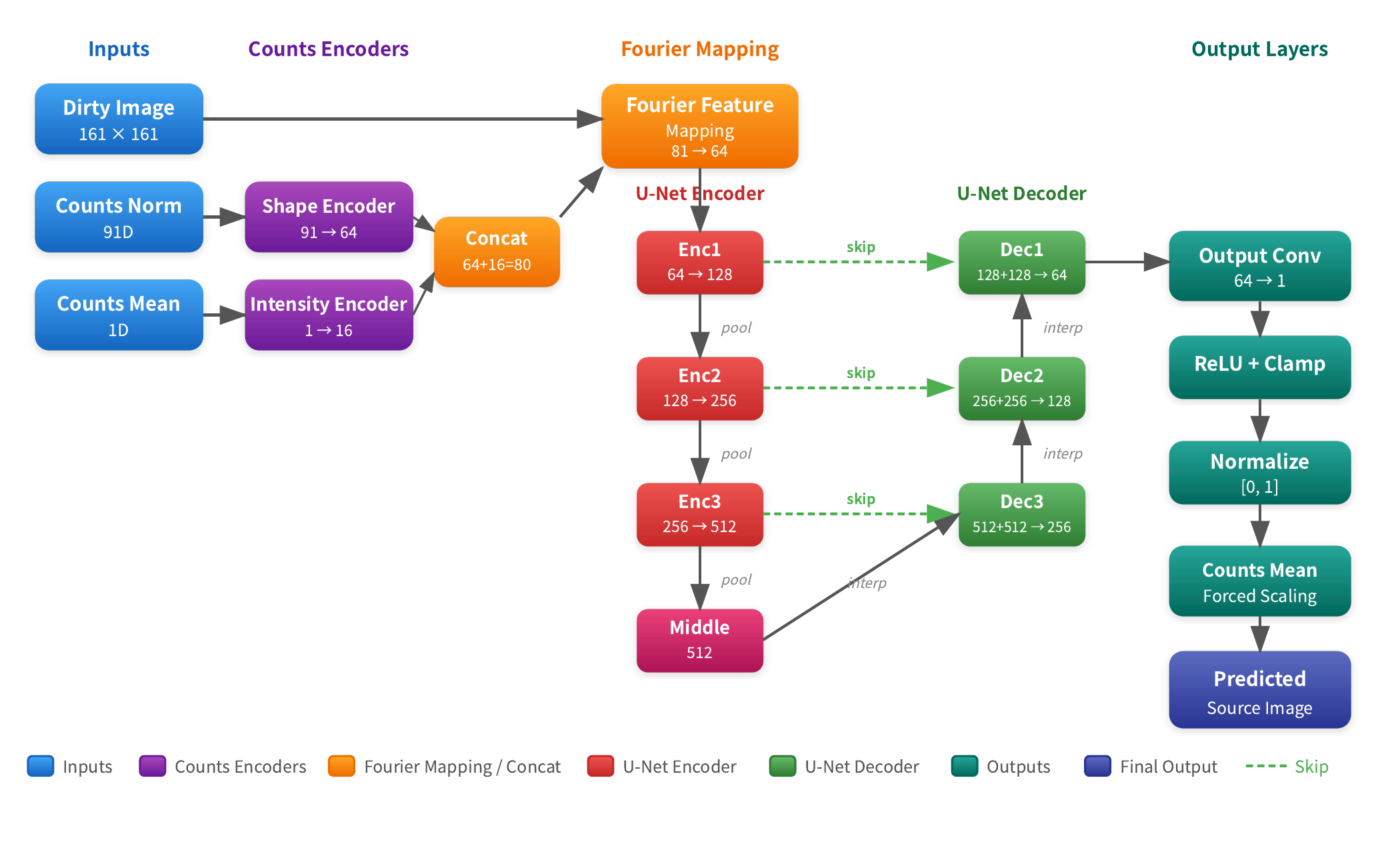}
\caption{Schematic of the HXI-DLA2 architecture. Inputs are the dirty image, the counts-distribution encoding (91$\to$64 dimensions), and the intensity encoding (1$\to$16 dimensions); the backbone combines multi-scale Fourier feature mapping with a three-level U-Net encoder--decoder; the output layer applies, in sequence, ReLU non-negative activation, clamping, peak normalization, and enforced counts-average-energy rescaling, guaranteeing the non-negativity and energy consistency of the reconstructed source image.}
\label{fig:3_network_architecture}
\end{figure*}

The overall network architecture is shown in Figure~\ref{fig:3_network_architecture}, comprising three modules: input design, backbone network, and output-layer physical constraints.

\textbf{Input design.} The network inputs contain two complementary types of physical information. The first is the dirty image, obtained by back-projecting the counts with the fixed patterns, which provides a low-resolution prior on the source position. Although the dirty image contains strong sidelobes and modulation fringes, its bright-region locations roughly correspond to the true source, serving as a spatial prior that narrows the network search space. The second is the 91-dimensional counts vector, decomposed via Equation~(\ref{eq:decouple}) into the average energy $\bar{C}$ (DC component, total source energy) and the distribution $\hat{\mathbf{C}}$ (AC component, spatial distribution), which are fed into an intensity encoder and a distribution encoder, respectively. The counts average energy $\bar{C}$ acts as an energy scalar constraint enforced through the forward mapping at the network output, guaranteeing that the total source energy of the reconstruction matches the observation; the counts distribution $\hat{\mathbf{C}}$ acts as a 91-dimensional spatial fingerprint enforced through a distribution-consistency term in the loss, so that the learned nonlinear mapping is trained to approximate the HXI physical response equation. The two encoded feature streams are concatenated with the dirty image along the channel dimension, forming a complete representation of the HXI physical measurement.

\textbf{Backbone network.} The backbone combines Fourier feature mapping \citep{Tancik2020} with a U-Net encoder--decoder \citep{Ronneberger2015}. Fourier features map coordinates into a high-frequency space, enabling the network to learn the multi-scale spatial-frequency responses associated with detectors of different pitches---this design directly corresponds to the wide spatial-frequency range of the 91 HXI detectors spanning 3\arcsec\ to 105\arcsec. The U-Net extracts multi-scale features through downsampling and preserves detail through skip connections, well-suited to complex multi-source flare structures. The counts-encoded features are broadcast and concatenated with the spatial feature maps, imposing physical counts guidance at every level of spatial learning and preventing the network from degenerating into a pure image-to-image translation model.

\textbf{Output-layer physical constraints.} The output layer predicts a single-channel source image and applies three layers of physical constraints in sequence. The first layer is the \textbf{non-negativity constraint}: a ReLU activation guarantees non-negativity, satisfying $I(x,y)\ge 0$, with an explicit \texttt{clamp} as a safeguard. Unlike the post-processing negativity truncation adopted in conventional pipelines---which changes the total image energy and violates energy conservation---our network embeds the ReLU in the end-to-end learning flow, adaptively learning to output non-negative images during training. The second layer is the \textbf{energy-consistency constraint}: enforced counts-average-energy rescaling is applied after the non-negativity truncation, aligning the average energy of the re-projected counts with the measured $\bar{C}$, guaranteeing that the total source energy remains consistent with the observation even after truncation. The third layer is the \textbf{distribution-consistency constraint}, implemented in the loss function (see Section~3.2), which drives the re-projected normalized counts distribution to match the measured $\hat{\mathbf{C}}$. The execution order of the three constraints---non-negativity truncation, average-energy rescaling, distribution supervision---ensures their hierarchical and non-violable nature.

\subsection{Physics-Inspired Loss Function}

The loss function is \textbf{dominated by counts-domain constraints, with image-domain constraints as auxiliary}.

The design motivations and synergistic relationships of the individual loss terms are as follows:

\textbf{Counts-distribution RMSE} is the core term, measuring the deviation of the normalized counts re-projected from the predicted image from the true counts distribution. The multi-channel weighting strategy assigns higher weights to high-frequency channels, because high-frequency channels correspond to small-pitch detectors that are more sensitive to fine source structure.

\textbf{Per-channel MAE} complements the global distribution RMSE with per-detector counts accuracy, preventing systematic biases in individual channels from being masked by overall averaging.

\textbf{Absolute counts closure} directly compares raw counts magnitudes to guarantee total source energy conservation. Together with the output-layer average-energy rescaling, it forms a dual safeguard: the output constraint ensures strict equality, while the closure term continuously guides the network to learn the correct energy scale during training.

\textbf{Image loss} serves only as an auxiliary to prevent the network from drifting away from a reasonable spatial position under counts constraints.

\textbf{Non-negativity penalty} works jointly with the output-layer ReLU activation, continuously guiding the network to produce non-negative source images during training.

The loss terms and their physical meanings are listed in Table~\ref{tab:loss_weights}.

\begin{table*}[ht!]
\centering
\caption{Components and physical meanings of the HXI-DLA2 loss function. The terms work jointly: the counts-related terms dominate the optimization and ensure that the reconstruction satisfies the HXI forward physical response, while the image loss and regularization terms provide auxiliary spatial-position constraints.}
\label{tab:loss_weights}
\footnotesize
\setlength{\tabcolsep}{6pt}
\begin{tabular}{lcc}
\toprule
Loss term & Physical meaning & Notes \\
\midrule
Counts-distribution RMSE & Normalized counts-distribution consistency & Frequency-weighted \\
Per-channel MAE & Per-detector counts accuracy & Frequency-weighted \\
Absolute counts closure & Total source energy conservation & Raw counts magnitude comparison \\
Image loss (L1/MSE) & Spatial position constraint & Full-image + dirty-mask auxiliary \\
Non-negativity penalty & Source image non-negativity & Joint with ReLU activation \\
\bottomrule
\end{tabular}
\end{table*}

\subsection{Dataset Construction and Training Strategy}

Two complementary source-image datasets are constructed for training and evaluation. The first is a Gaussian simulated dataset: source images are generated analytically with controllable parameters (position, standard deviation, peak intensity), and the corresponding 91-dimensional counts and labels are computed exactly through the forward model (Equation~\ref{eq:forward}). With exact ground truth and fully controlled parameters, this dataset provides the controlled benchmarks for the Gaussian morphology experiments (Section~3.4), the double-source dynamic-range experiments (Section~3.5), and the extreme-ratio experiment (Section~5.1). The second is a soft X-ray dataset built from Hinode/XRT observations with the Be-thick (119 $\mu$m) filter, covering real flare morphologies---single bright cores, loop-like structures, compact cores, and diffuse structures; it provides the main training corpus and the complex-morphology test cases evaluated in Section~4.1, and the resulting network is then applied to a real HXI observation event in Section~4.2.

Since modulation-based hard X-ray imaging has no direct-imaging reference data, soft X-ray images---currently the only abundant observations that reflect the true spatial structures of flares---are adopted as training labels; the practical advantages and inherent limitations of this choice are analyzed in Section~4.1. Although soft X-ray emission and the nonthermal hard X-ray emission targeted by HXI differ in mechanism and morphology, every label is modulated through the fixed patterns according to Equation~(\ref{eq:forward}) to generate the counts: what the network learns is the mapping between the counts distribution and the forward physical response, which is uniquely determined by the instrumental patterns and independent of the energy band of the source images. A small number of Gaussian simulated sources are additionally mixed into the training set, keeping the network anchored to basic physical laws rather than only the statistical biases of particular morphologies.

In HXI imaging, each energy band corresponds to an independent set of 91 modulation patterns determined by in-flight calibration, because the grid transmission and detector response vary with photon energy. The experiments in this paper use the modulation patterns at the 20 keV energy band. All source images are $161\times161$ pixels. The training and validation sets use Hinode/XRT observation events from 2012 January to 2024 May, split in an 8:1 ratio; the test set uses observation events from 2024 June to 2024 December, \textbf{completely independent} of the training and validation phases. Training uses the Adam optimizer \citep{Kingma2015} with an initial learning rate of $10^{-4}$ and a batch size of 32 on an NVIDIA H100 GPU, converging in about 7 days, with early stopping preserving the best model.

\subsection{Gaussian Simulated Source Experiments}

The three morphologies are chosen to match the validation goals: the single source probes baseline reconstruction fidelity, while the double source (1:5 intensity ratio) and the ring source represent the multi-component and extended configurations on which the purely data-driven reconstruction of our earlier HXI-DLA study degraded; dedicated double-source dynamic-range experiments follow in Section~3.5. The test sources adopt the same parameters as in the earlier HXI-DLA benchmark, ensuring direct continuity with our group's previous experiment, and the input counts and labels are computed analytically from the forward model (Equation~\ref{eq:forward}).

\begin{figure*}[ht!]
\centering
\includegraphics[width=0.9\textwidth]{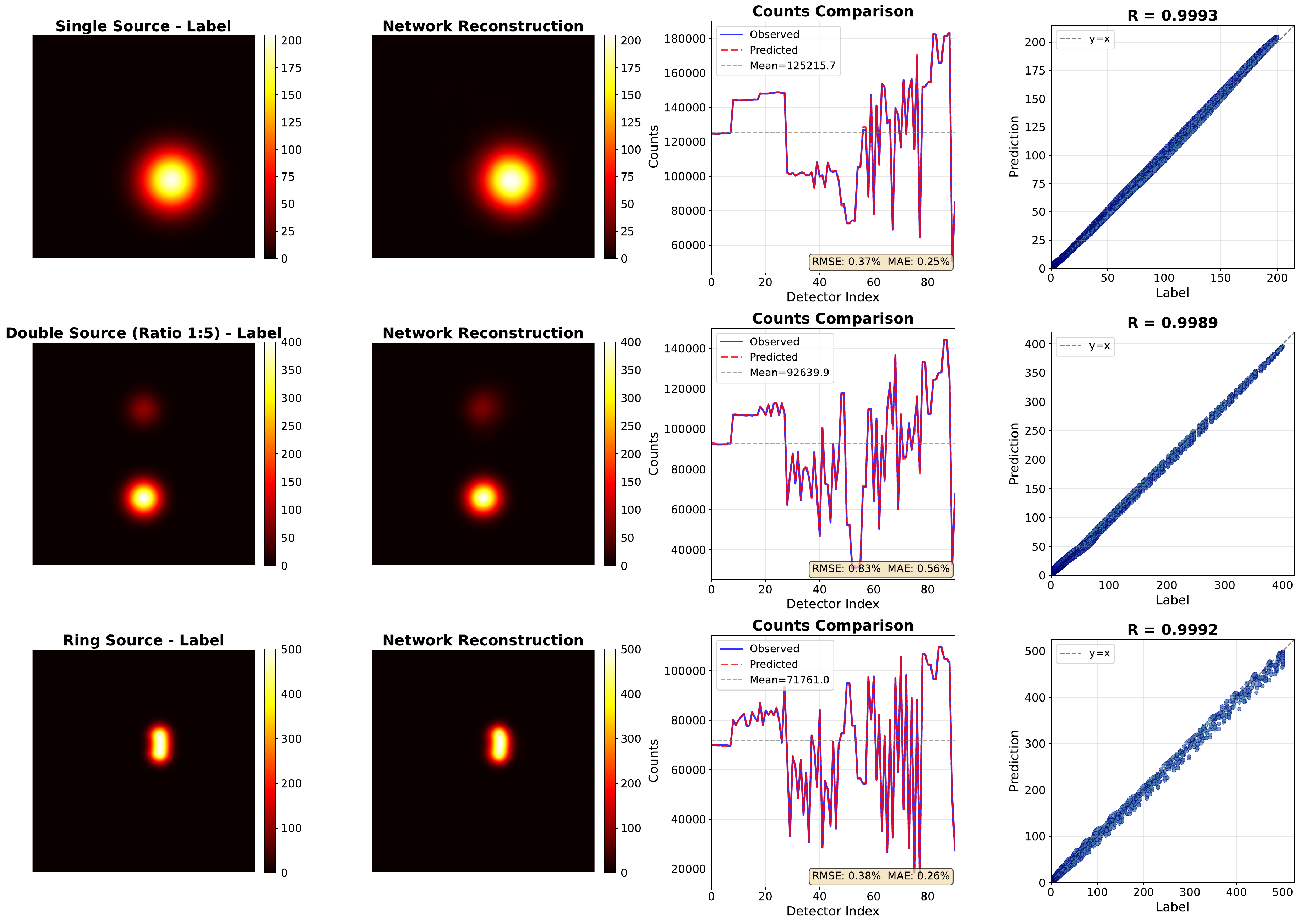}
\caption{HXI-DLA2 reconstruction results on three Gaussian source morphologies. From top to bottom: single source, double source (1:5 intensity ratio), and ring source. Each row shows, from left to right: the ground-truth label, the network reconstruction, the normalized counts comparison between the label and the reconstruction, and the pixel-wise prediction--label correlation scatter plot (annotated with the correlation coefficient $R$).}
\label{fig:4_1_source_types}
\end{figure*}

The reconstructions are shown in Figure~\ref{fig:4_1_source_types}, with quantitative metrics summarized in Table~\ref{tab:gauss_compare}; the image-domain chi-square reported there is $\chi^2=\sum_{x,y}[I_{\rm rec}(x,y)-I_{\rm true}(x,y)]^{2}/I_{\rm true}(x,y)$, computed pixel by pixel between the reconstruction and the reference image over the pixels with $I_{\rm true}(x,y)>0$. Two characteristics of the results stand out. First, the image-domain fidelity is uniformly high ($R\gtrsim0.999$, SSIM~$\ge0.98$, $\chi^2<4$), and the residual errors grow only mildly with morphological complexity, a trend examined together with the other experiments in Section~5.1. Second, the counts panels show that the re-projected normalized counts of the reconstructions nearly coincide with the labels, while the per-channel MAE (Table~\ref{tab:gauss_compare}) increases with source complexity. In particular, the double and ring sources---on which the purely data-driven HXI-DLA degraded---are reconstructed faithfully here. This improvement does not come from a larger model but from the constraint structure: HXI-DLA learned only a counts-to-image mapping under pixel-level supervision, so on multi-component or extended sources, where the counts-to-image mapping is most non-unique, it could converge to solutions inconsistent with the forward equation. HXI-DLA2 instead confines the output to the forward-feasible subspace---non-negativity and counts-average-energy closure at the output, and a distribution-consistency loss tying the morphology to the counts phase signature---so the degenerate solutions responsible for the earlier degradation are excluded by construction.

\begin{table*}[ht!]
\centering
\caption{Quantitative results of HXI-DLA2 reconstruction on three Gaussian source morphologies. Image-domain metrics: NRMSE (normalized root-mean-square error), $\chi^2$ (chi-square statistic), SSIM (structural similarity index), $R$ (Pearson correlation coefficient); counts-domain metric: MAE (per-channel mean absolute error, in counts); composite index: QuIX (a comprehensive quality index for hard X-ray image reconstruction combining pixel accuracy, structure preservation, and sidelobe suppression).}
\label{tab:gauss_compare}
\footnotesize
\setlength{\tabcolsep}{4pt}
\begin{tabular}{lcccccc}
\toprule
Source Type & NRMSE & MAE & $\chi^2$ & SSIM & $R$ & QuIX \\
\midrule
Single Source        & 0.00581 & 189.9 & 0.76 & 0.9824 & 0.9993 & 0.9899 \\
Double Source (1:5)  & 0.00456 & 313.9 & 3.45 & 0.9862 & 0.9989 & 0.9920 \\
Ring Source          & 0.00355 & 517.5 & 1.72 & 0.9979 & 0.9992 & 0.9976 \\
\bottomrule
\end{tabular}
\end{table*}

\subsection{Double-Source Dynamic Range Experiments}

The above experiments include only a double source with a small intensity contrast (1:5), and the wide dynamic range between strong and weak sources has not been tested. We therefore proceed to double-source dynamic-range experiments.

\begin{figure*}[ht!]
\centering
\includegraphics[width=0.88\textwidth]{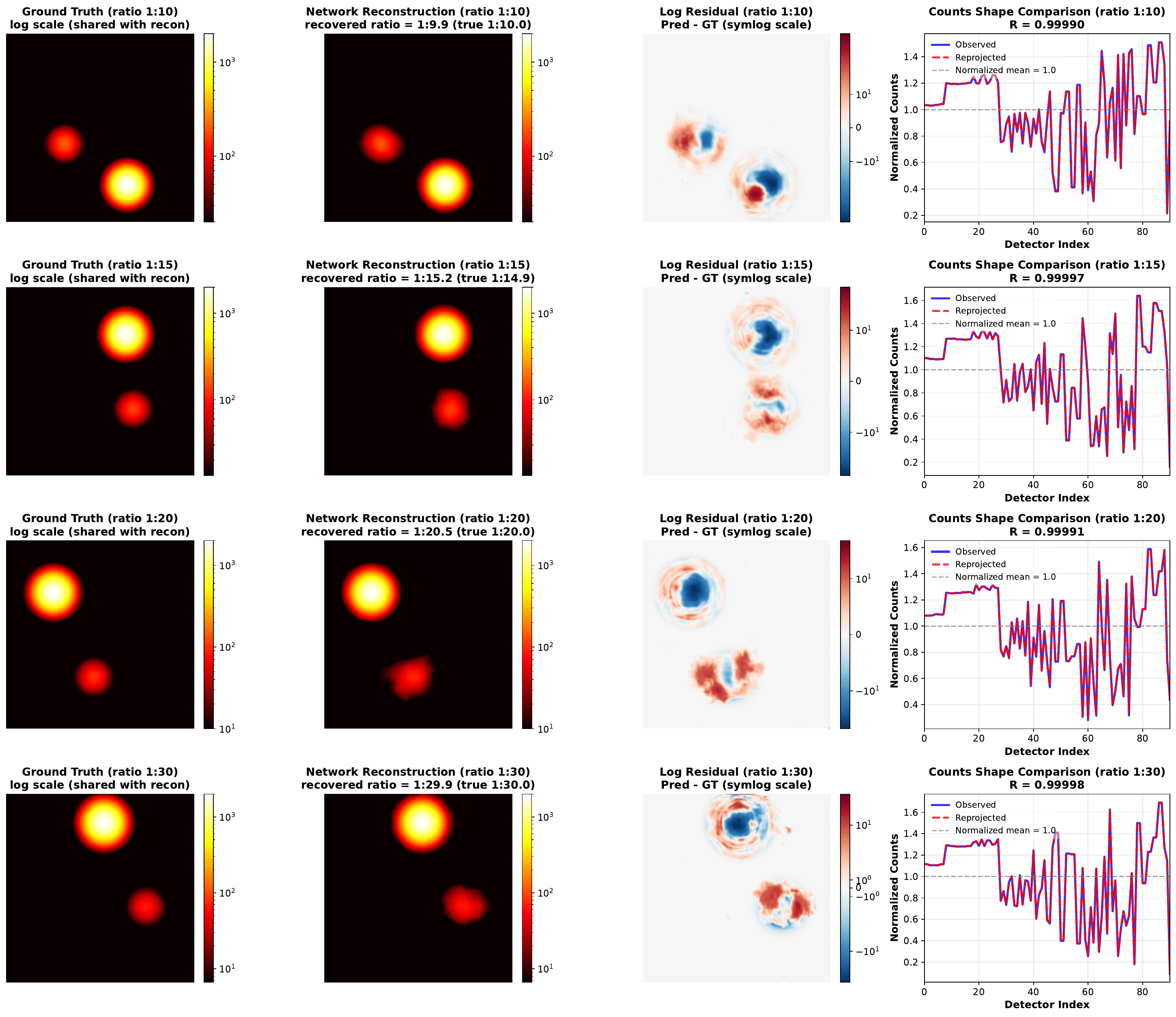}
\caption{Results of the double-source dynamic-range experiments. Each row corresponds to one peak intensity ratio (from top to bottom: 1:10, 1:15, 1:20, and 1:30). The four columns show, from left to right: the ground-truth label on a logarithmic scale, the network reconstruction on a logarithmic scale (annotated with the recovered peak ratio), the logarithmic residual map (prediction minus label, symlog color scale), and the counts-distribution comparison (observed versus re-projected normalized counts with the correlation coefficient). In all four experiments the counts of the predicted and target source images agree, with counts-distribution correlation coefficients above 0.9999.}
\label{fig:4_1_two_source}
\end{figure*}

Double-source reconstruction is a challenge in modulation-based imaging inversion: the counts response is the superposition of two sub-source modulations, and the network must recover the spatial position and intensity ratio of each source from the coupled counts. When the weak source is much fainter than the strong one, its contribution to the counts is masked, and whether it can be resolved depends directly on the network's sensitivity to subtle counts differences. Benchmark studies quantify how difficult this is in practice: \citet{Yu2025} tested the imaging dynamic range of HXI with double sources and found that, for extended Gaussian sources, the weak source already falls below the sidelobe level near a ratio of 20:1 and becomes completely unresolved at 30:1 with HXI\_CLEAN, corresponding to a dynamic range of roughly 10--20:1; only point sources remain discernible up to about 50:1. In the earlier HXI-DLA tests, the double-source dynamic range reached only about 5:1 \citep{Xia2024}. Both studies assess detectability through the flux or peak-intensity ratio of the two sources, and we adopt the same peak-ratio criterion here. Building on that work, this section performs in-depth double-source dynamic-range experiments with Gaussian simulated sources, varying both the spatial positions and the peak intensity ratios of the two sources, covering peak ratios of 1:10, 1:15, 1:20, and 1:30. The evaluation combines image-domain metrics---NRMSE, SSIM, total-flux relative error, per-source peak and centroid errors, and the reconstruction--label correlation $R$ (Table~\ref{tab:two_source_results})---with a measurement-domain comparison between the observed and re-projected counts profiles (Figure~\ref{fig:4_1_two_source}).

The results are shown in Figure~\ref{fig:4_1_two_source} and summarized in Table~\ref{tab:two_source_results}. Three trends are evident. First, both sources are detected at every ratio from 1:10 to 1:30, the counts average energy is closed exactly, and the re-projected normalized counts remain close to the input; the global image-domain metrics are high and almost insensitive to the intensity ratio (NRMSE~$\sim10^{-3}$, SSIM~$>0.99$, $R>0.9999$), indicating that the reconstruction quality is governed by the morphology rather than by the intensity contrast. Second, the residual errors are asymmetric between the two sources: the strong source is recovered almost exactly (peak error $\lesssim$2\%, centroid error $\lesssim$0.25~px), whereas the weak-source errors grow with the ratio---its peak is underestimated by up to $\sim$4.5\% and its centroid error approaches one pixel at 1:30, since a weak source embedded in the sidelobe field of the strong source is the component most easily displaced by modulation fringes. Third, because both sources remain reliably resolved at 1:30, the resolution limit has not yet been reached within the tested range.

\begin{table*}[ht!]
\centering
\caption{Image-domain and measurement-domain quantitative results of the double-source dynamic-range experiments. NRMSE is the normalized root-mean-square error, SSIM the structural similarity index, Flux RE the relative error of the total source flux, Peak RE the peak relative error (+ overestimate, $-$ underestimate), Cent the centroid position error (pixels), and $R$ the Pearson correlation coefficient between the reconstructed and label images.}
\label{tab:two_source_results}
\footnotesize
\setlength{\tabcolsep}{5pt}
\begin{tabular}{ccccccccc}
\toprule
Ratio & NRMSE & SSIM & Flux RE & Strong Peak RE & Weak Peak RE & Strong Cent & Weak Cent & $R$ \\
\midrule
1:10 & 0.0027 & 0.9962 & +0.02\%  & $-$1.00\% & $-$0.49\% & 0.22\,px & 0.11\,px & 0.99990 \\
1:15 & 0.0015 & 0.9980 & $-$0.07\% & $-$0.53\% & $-$2.61\% & 0.08\,px & 0.34\,px & 0.99997 \\
1:20 & 0.0023 & 0.9949 & $-$0.10\% & $-$1.98\% & $-$4.49\% & 0.12\,px & 0.27\,px & 0.99991 \\
1:30 & 0.0016 & 0.9977 & $-$0.06\% & $-$1.74\% & $-$1.45\% & 0.03\,px & 1.00\,px & 0.99998 \\
\bottomrule
\end{tabular}
\end{table*}

Beyond the 1:30 dynamic range, we further push to the extreme limit of 1:35--1:50; the results and analysis are presented in the Discussion (Section~5.1).

\section{Complex Source Reconstruction}

This section validates the reconstruction capability and physical consistency of HXI-DLA2 at two levels: real complex morphologies from Hinode/XRT soft X-ray observations for generalization testing (Section~4.1), and a real HXI flare event validated against contemporaneous SDO/AIA multi-band observations (Section~4.2). Quantitative evaluation employs image-domain metrics (NRMSE, SSIM, $R$, $\chi^2$), counts-domain metrics (MAE), and the QuIX composite index \citep{Li2025}, a comprehensive evaluation metric for hard X-ray imaging algorithms that combines pixel accuracy, structure preservation, and sidelobe suppression into a single scalar.

\subsection{Soft X-Ray Reconstruction Experiments}

To assess the reconstruction of complex spatial structures, this section uses Hinode/XRT soft X-ray source images (Be-thick filter, 1.5--2.5 keV) as test labels, drawn from the temporally independent test set described in Section~3.3. Soft X-ray observations offer practical advantages for dataset construction: XRT provides a large, well-calibrated archive of high-resolution, high signal-to-noise flare images spanning diverse morphologies, and because soft X-ray emission outlines the heated flare-loop plasma, its spatial distribution serves as a physically plausible proxy for the source structures that hard X-ray imaging aims to recover. The choice also has an inherent limitation: soft X-ray morphology is dominated by thermal loop emission, whereas nonthermal hard X-ray sources are often more compact and footpoint-like. This experiment therefore evaluates the spatial reconstruction capability on complex thermal morphologies and does not by itself demonstrate the correctness of nonthermal HXR morphology recovery.

\begin{figure*}[ht!]
\centering
\includegraphics[width=0.85\textwidth]{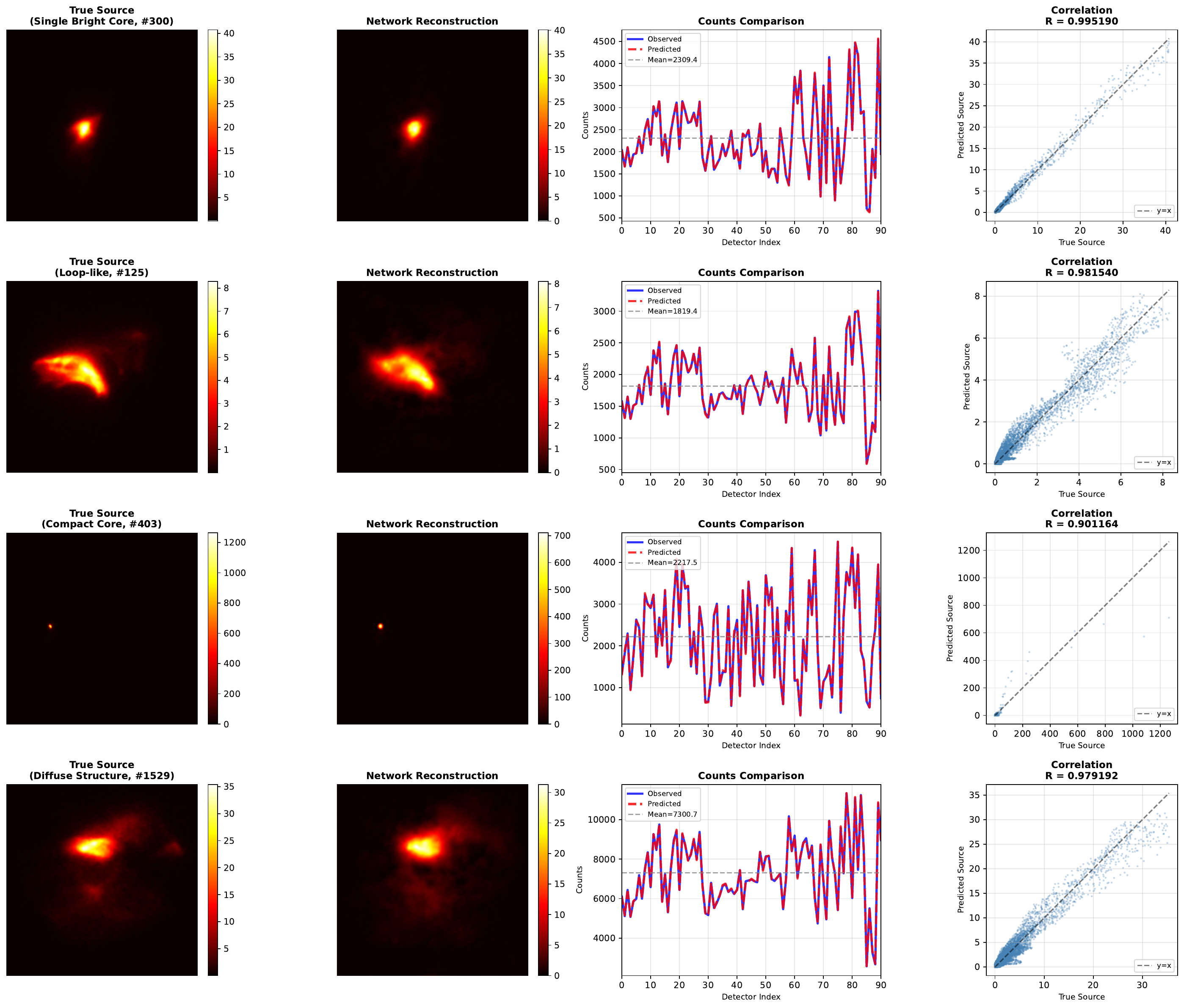}
\caption{Representative soft X-ray source morphologies in the training set. The four rows show, from top to bottom: single bright core, loop-like structure, compact core, and diffuse structure; each row displays, from left to right: the label source image, the network reconstruction, the normalized counts curve comparison, and the pixel-wise label--reconstruction correlation scatter plot (annotated with $R$). The training set essentially covers the common spatial structure types of solar flare soft X-ray sources.}
\label{fig:4_2_morphologies}
\end{figure*}

\textbf{Morphological diversity of the training set.} As described in Section~3.3, the training set covers four typical flare morphologies, with representative examples shown in Figure~\ref{fig:4_2_morphologies} and their quantitative metrics listed in the training-set rows of Table~\ref{tab:sxr_complex_results}. In all cases the re-projected counts already agree closely with the input, yet the image-domain reconstruction--label correlation is not perfect and the compact-core peak deviates visibly---counts agreement alone does not uniquely determine the image, which is the intrinsic limitation of this underdetermined inversion.

\begin{figure*}[ht!]
\centering
\includegraphics[width=0.88\textwidth]{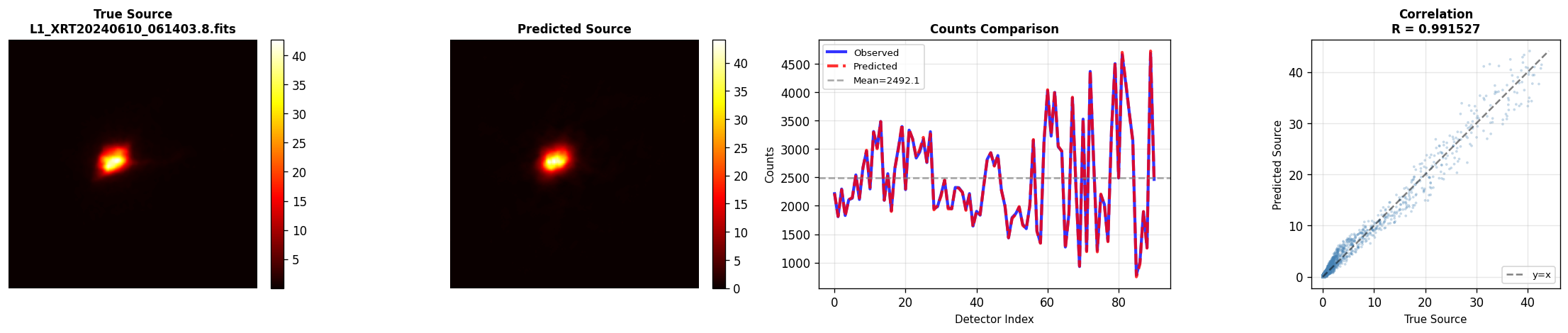}
\\[4pt]
\includegraphics[width=0.88\textwidth]{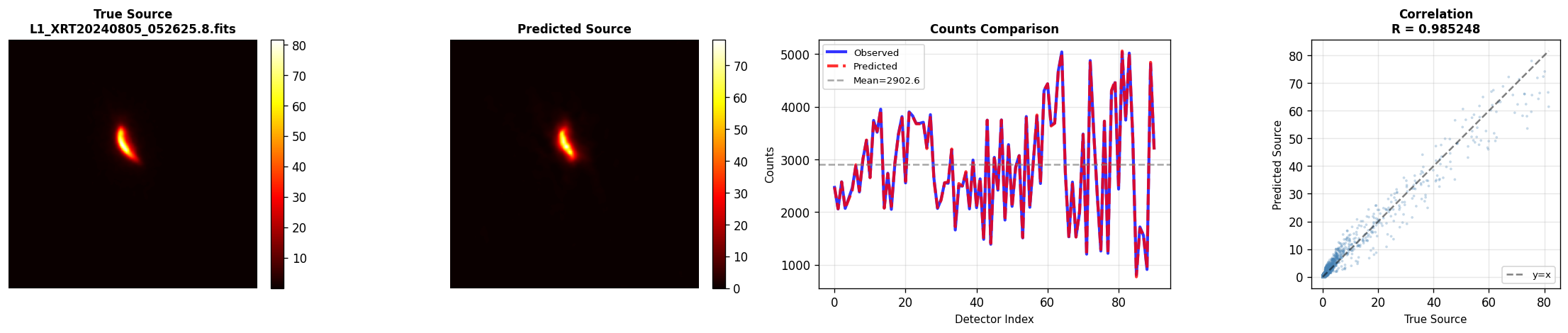}
\\[4pt]
\includegraphics[width=0.88\textwidth]{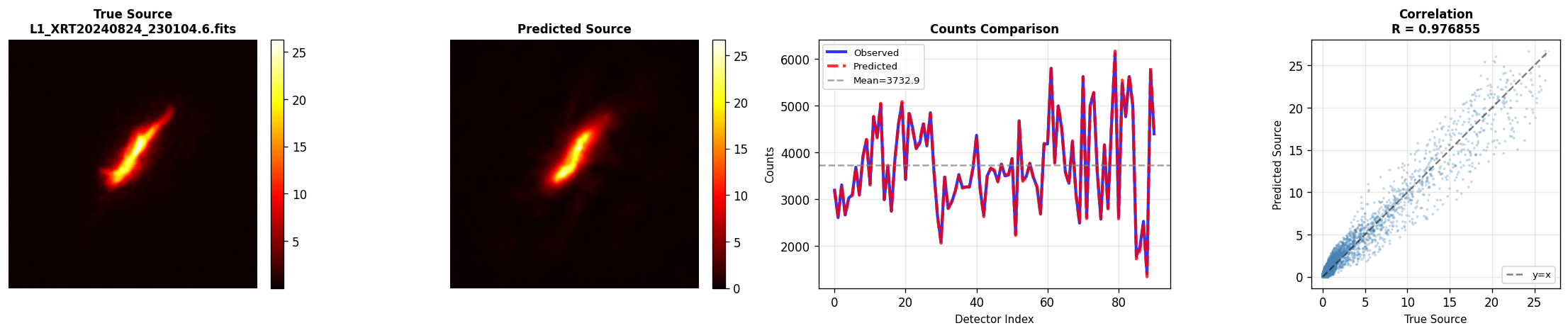}
\caption{Reconstruction of soft X-ray observation events and physical-consistency verification. The three rows correspond to Hinode/XRT Be-thick filter observations on 2024 June 10 (event XRT20240610\_061403.8), August 5 (event XRT20240805\_052625.8), and August 24 (event XRT20240824\_230104.6). In each group, columns from left to right: original XRT source image, network reconstruction, normalized counts-distribution comparison (observed versus predicted curves), and the pixel-wise correlation scatter between the true and predicted source images (annotated with the image-domain correlation coefficient $R$).}
\label{fig:4_2_complex}
\end{figure*}

\textbf{Reconstruction results.} Three real observation events (2024 June 10, August 5, and August 24) are selected from the temporally independent test set for reconstruction validation. Because the test events are temporally separated from the training data, they probe the model's generalization to unseen observations. Figure~\ref{fig:4_2_complex} shows the reconstructions and physical-consistency verification for the three observation events, with quantitative results summarized in Table~\ref{tab:sxr_complex_results}. Event XRT20240610\_061403.8 (first row) has a compact kernel structure; the network accurately recovers the kernel position and peak intensity. Event XRT20240805\_052625.8 (second row) has a small loop-like structure with a low peak; the network still accurately recovers the loop morphology and source position. Event XRT20240824\_230104.6 (third row) has a high-brightness loop-like structure; the network recovers the overall loop contour and main bright-region position, but the peak is somewhat underestimated and fine-scale structure is smoothed. Quantitative metrics for the three events are summarized in Table~\ref{tab:sxr_complex_results}.

In all three tests the predicted counts average energy matches the input exactly. The peak deviation in the third event is relatively large because the source image has a wide brightness distribution with a high proportion of loop-like structure; the 91-dimensional counts provide weaker pixel-level constraints on such high-dynamic-range loop-like sources, and the network output applies prior regularization to the peak while maintaining counts consistency.

\begin{table*}[ht!]
\centering
\caption{Quantitative results of soft X-ray complex-morphology reconstruction on four representative training-set morphologies and three temporally independent test events. Image-domain metrics: NRMSE, $\chi^2$, SSIM, $R$; counts-domain metric: MAE (per-channel mean absolute error, in counts); composite index: QuIX, a comprehensive quality index for hard X-ray image reconstruction combining pixel accuracy, structure preservation, and sidelobe suppression.}
\label{tab:sxr_complex_results}
\footnotesize
\setlength{\tabcolsep}{4pt}
\begin{tabular}{lcccccc}
\toprule
Sample & NRMSE & MAE & $\chi^2$ & SSIM & $R$ & QuIX \\
\midrule
\multicolumn{7}{l}{\emph{Training set (Figure~\ref{fig:4_2_morphologies})}} \\
Single Bright Core & 0.0058 & 8.666 & 0.0628 & 0.9870 & 0.9952 & 0.9952 \\
Loop-like & 0.0224 & 8.467 & 0.0692 & 0.9070 & 0.9815 & 0.9814 \\
Compact Core & 0.0045 & 20.382 & 0.3993 & 0.9990 & 0.9012 & 0.8837 \\
Diffuse Structure & 0.0189 & 3.360 & 0.0250 & 0.8923 & 0.9792 & 0.9783 \\
\midrule
\multicolumn{7}{l}{\emph{Test set (Figure~\ref{fig:4_2_complex})}} \\
XRT20240610\_061403.8 & 0.0081 & 1.480 & 0.0154 & 0.9795 & 0.9915 & 0.9915 \\
XRT20240805\_052625.8 & 0.0080 & 17.058 & 0.2068 & 0.9864 & 0.9852 & 0.9839 \\
XRT20240824\_230104.6 & 0.0187 & 2.269 & 0.0263 & 0.8954 & 0.9769 & 0.9767 \\
\bottomrule
\end{tabular}
\end{table*}

\subsection{Validation on Real HXI Observations}

\textbf{Observation overview.} We validate the method on a flare observed by ASO-S/HXI on 2024 March 23 (13:50:52.626--13:51:08.626 UTC). This event was selected because it exhibits a relatively high hard X-ray signal-to-noise ratio within the HXI observation window, and the observational coverage of ASO-S and SDO/AIA overlaps well, facilitating multi-band cross-validation. The observation covers the hard X-ray enhancement around the flare peak with 64 accumulated exposures totaling about 16 s, centered at solar coordinates $(-100'', -120'')$, with a pixel scale of $1''$ per pixel. Each exposure records the raw counts of the 91 detectors; after dark-field and gain correction, the counts are accumulated over time to yield raw counts (mean 1548.89, standard deviation 519.43).

\begin{figure*}[ht!]
\centering
\includegraphics[width=0.75\textwidth]{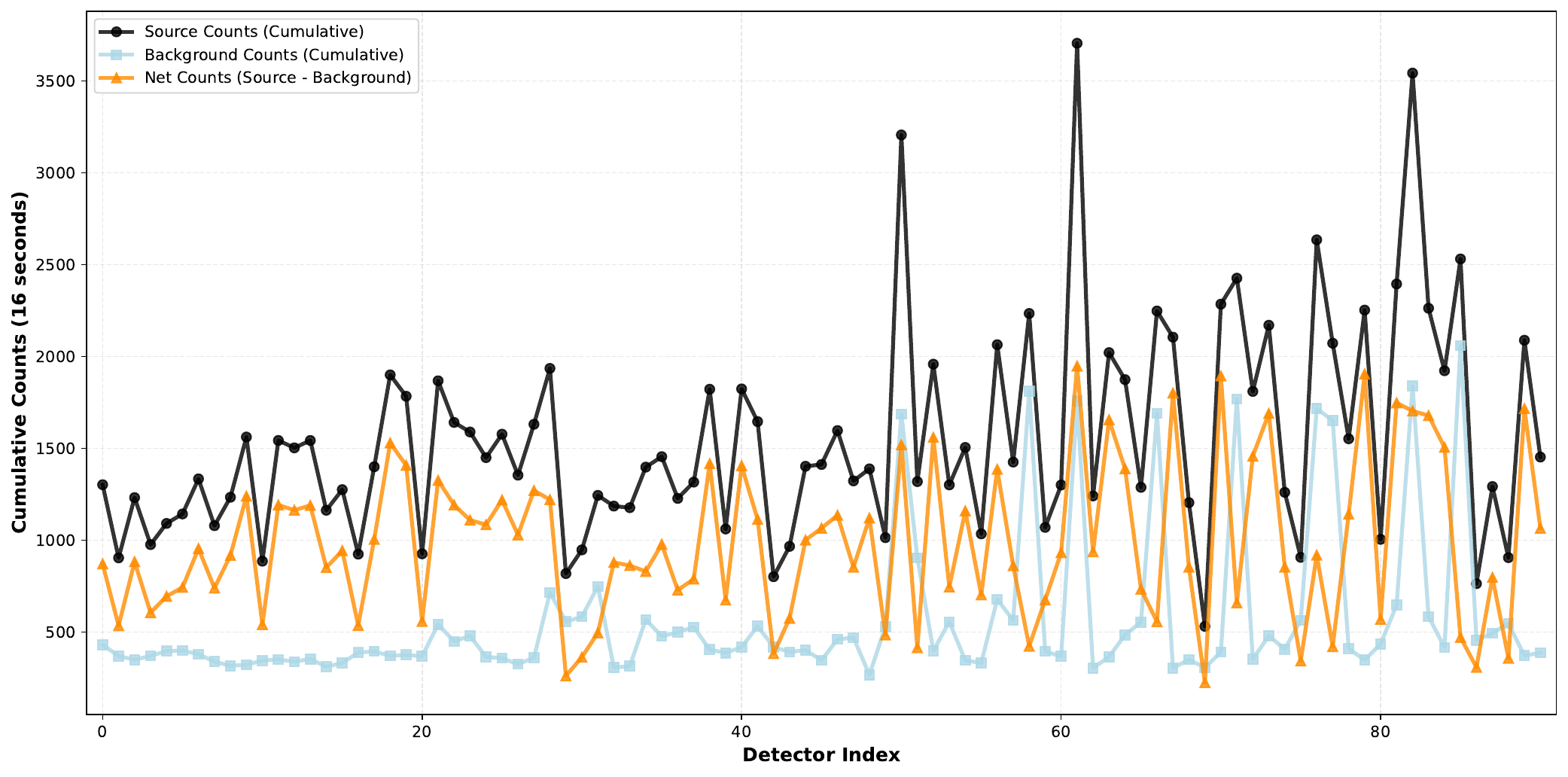}
\caption{Accumulated counts of the 91 detectors for the 2024 March 23 flare (13:50:52.626--13:51:08.626 UTC, 64 exposures). Black: raw counts; light blue: background counts; orange: net counts, which serve as the network input.}
\label{fig:4_3_counts_curve}
\end{figure*}

Background estimation follows the standard off-flare time window approach: detector counts are accumulated over observation windows approximately 48 hr before and after the flare event, when no flare signal is present, and the per-channel mean is taken as the background estimate (mean 563.71, standard deviation 417.24). This method assumes that the non-flare background is temporally stable, but in practice the background is affected by cosmic rays, ambient particle radiation, and detector dark current, leading to per-channel fluctuations. Subtracting the background estimate from the raw counts yields net counts (mean 985.18, standard deviation 426.84). The background per-channel fluctuations are comparable to the background mean itself (std/mean ratio $\sim$74\%), indicating substantial uncertainty in the background estimate---under- or over-subtraction directly alters the average energy and distribution of the net counts, thereby affecting the energy scale and spatial morphology of the reconstruction. Figure~\ref{fig:4_3_counts_curve} shows the raw, background, and net counts of the 91 detectors. The dirty image back-projected from the net counts serves as the network input, with a peak of about 2412 and strong negative sidelobes and modulation fringes.

\begin{figure*}[ht!]
\centering
\includegraphics[width=0.95\textwidth]{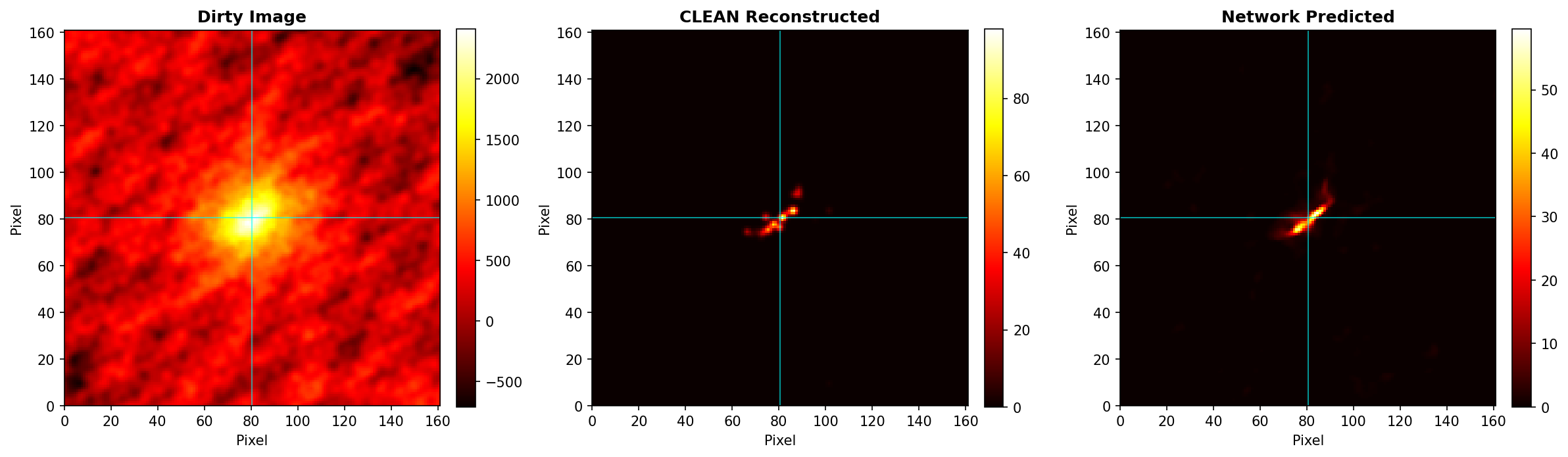} \\[2pt]
\small (a) Image-domain comparison: the back-projected dirty image (network input, left), the standard HXI data product (CLEAN, middle), and the HXI-DLA2 reconstruction (right) \\[4pt]
\includegraphics[width=0.72\textwidth]{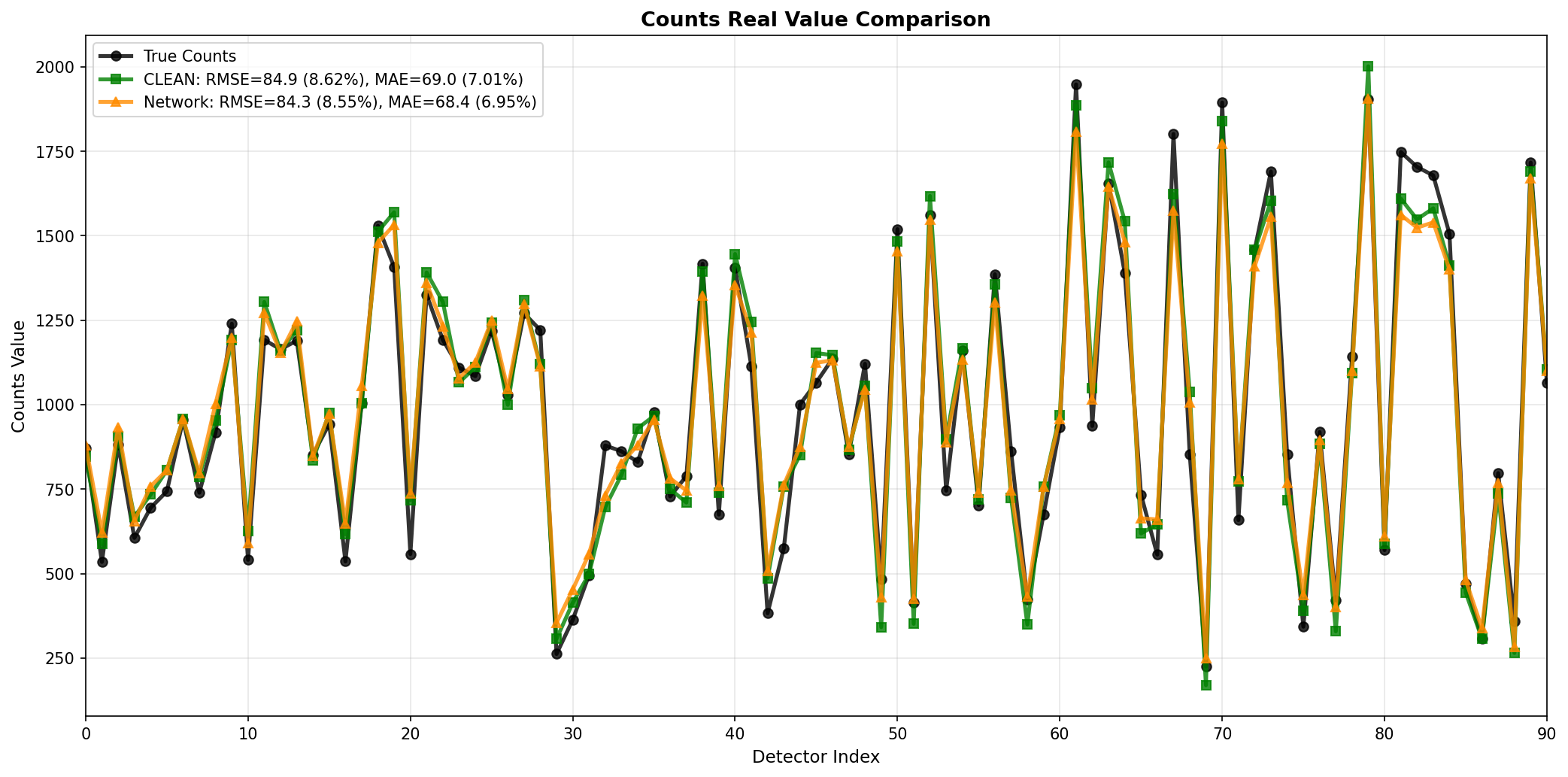} \\[2pt]
\small (b) Measurement-domain verification: observed net counts (black), counts re-projected from the HXI-DLA2 reconstruction (orange), and counts re-projected from the standard data product (green, shown for reference)
\caption{Reconstruction of the real HXI observation event, shown jointly in the image and measurement domains. (a) The dirty image back-projected from the net counts (the network input, left), the standard HXI data product (CLEAN, middle) as the conventional reference, and the HXI-DLA2 reconstruction (right), which yields a continuous, smooth source morphology through end-to-end physical constraints. (b) Observed net counts (black) versus counts re-projected from the HXI-DLA2 reconstruction (orange), achieving RMSE 8.55\% and MAE 6.95\%; the re-projection of the standard data product (green) is overplotted as the conventional reference. Together the two panels connect the reconstructed image morphology with its consistency with the observed counts.}
\label{fig:4_network_vs_clean}
\end{figure*}

\textbf{Reconstruction result.} Feeding the dirty image and net counts into the trained network directly yields the reconstructed source image. In conventional iterative deconvolution pipelines, the reconstruction depends on manually tuned parameters---such as the number of iterations, the loop gain, and the beam width---and different parameter combinations may produce different results. In contrast, the HXI-DLA2 network output is deterministic: given the same dirty image and net counts, the network always produces a unique reconstruction without any manual parameter choices. The dirty image serving as the network input, the standard HXI CLEAN data product as the conventional reference, and the resulting network reconstruction are compared together in Figure~\ref{fig:4_network_vs_clean}(a).

\textbf{Quantitative analysis.} Figure~\ref{fig:4_network_vs_clean} presents the reconstruction at two complementary levels: panel (a) compares the three stages of the inversion in the image domain---the dirty image back-projected from the net counts (the network input), the standard HXI data product (CLEAN, the conventional reference), and the network output, which yields a continuous and smooth source morphology without any post-processing truncation; panel (b) verifies the consistency in the measurement domain, comparing the observed net counts with the counts re-projected from the reconstruction and from the standard data product. At the physical-consistency level, the predicted counts average energy agrees with the measured value to within numerical precision, so the enforced closure is preserved under real noise. The counts re-projected from the network reconstruction achieve an RMSE of 8.55\% and an MAE of 6.95\% against the observed net counts; this error level is markedly higher than on simulations (RMSE $<1$\%), for reasons discussed in Section~5.1. The network output peak is of the same order as that of the standard data product, consistent with the enforced average-energy rescaling. Whereas conventional pipelines obtain a visually clear source region only through post-hoc negativity truncation, HXI-DLA2 satisfies non-negativity and energy conservation intrinsically through end-to-end optimization (ReLU activation and average-energy rescaling), yielding a natural source morphology without post-processing.

\begin{figure*}[ht!]
\centering
\includegraphics[width=0.95\textwidth]{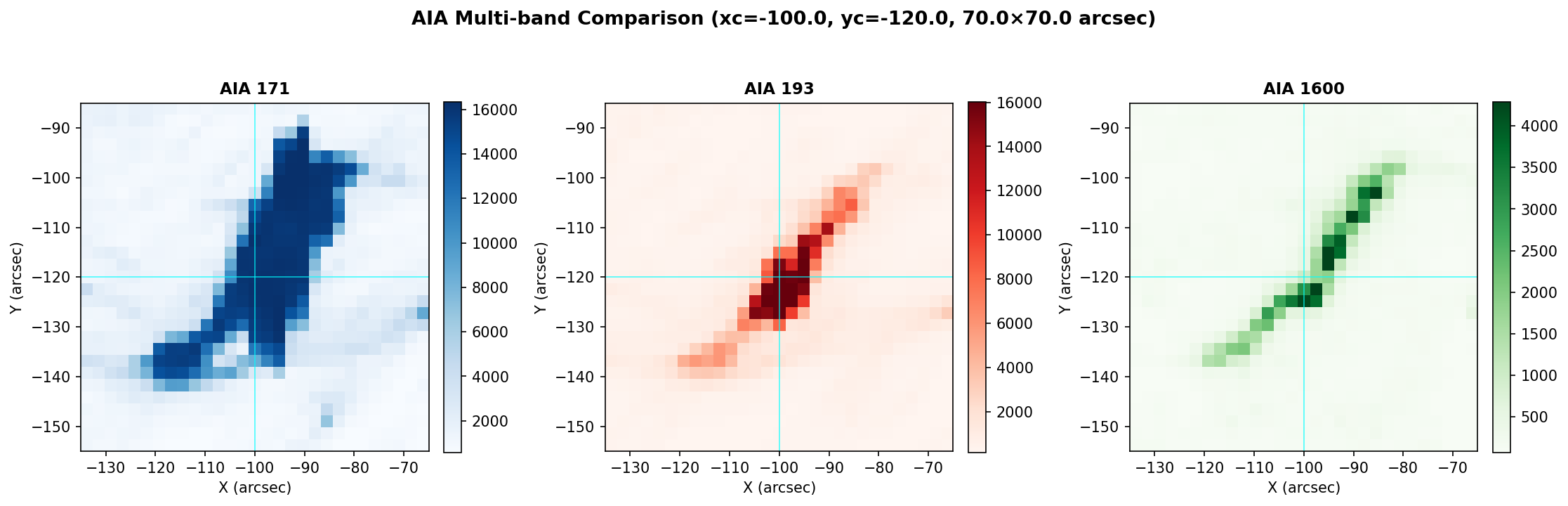} \\[2pt]
\small (a) Contemporaneous SDO/AIA observations at 171~\AA, 193~\AA, and 1600~\AA\ (field of view matching the HXI reconstruction region) \\[4pt]
\includegraphics[width=0.65\textwidth]{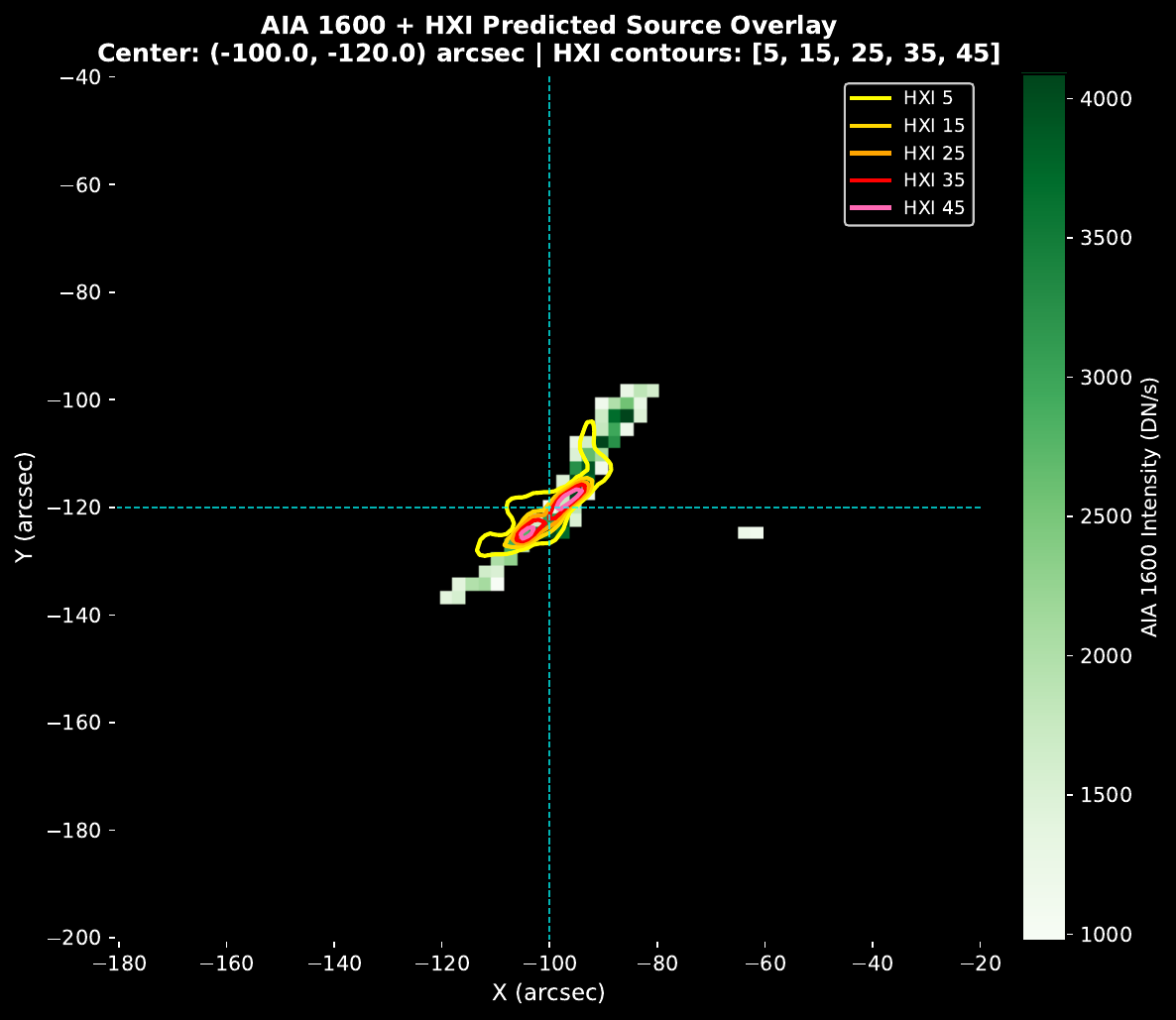} \\[2pt]
\small (b) Contours of the HXI-DLA2 reconstruction overlaid on the AIA 1600~\AA\ image
\caption{Multi-band morphological validation of the real-event reconstruction. (a) Contemporaneous flare loop structures observed by SDO/AIA at 171~\AA, 193~\AA, and 1600~\AA\ (1024$\times$1024 images binned by 4, pixel scale 2.4\arcsec; field of view $70''\times70''$ centered at $(-100'', -120'')$, matching the HXI reconstruction region), serving as an independent reference for the spatial morphology. (b) Contours of the HXI-DLA2 reconstruction overlaid on the AIA 1600~\AA\ image: the hard X-ray bright region agrees closely with the footpoint locations and extension direction of the AIA loop structure; the two bands probe plasma emission at different temperatures and are not expected to correspond exactly, but the spatial distributions of the main energy-release regions are consistent.}
\label{fig:4_network_vs_aia}
\end{figure*}

\textbf{Multi-band spatial comparison.} Contemporaneous SDO/AIA observations provide an independent spatial reference for the flare loop structure (Figure~\ref{fig:4_network_vs_aia}(a)): the 171~\AA\ band primarily reflects coronal loop plasma at $\sim$0.8~MK, 193~\AA\ traces hotter loop structures at $\sim$1.5~MK, and 1600~\AA\ probes the transition region and chromospheric UV emission; together, the three bands constrain the three-dimensional spatial structure of the flare loop. Compared with the AIA 1600~\AA\ observation (Figure~\ref{fig:4_network_vs_aia}(b)), the hard X-ray bright region aligns closely with the footpoint locations and extension direction of the AIA loop structure, verifying the overall spatial consistency between the reconstructed bright region and the known flare loop structure. The overlaid contours of the reconstruction show that the main body of the hard X-ray bright region falls near the footpoints of the AIA 1600~\AA\ loop structure, with the extension direction consistent with the loop axis, indicating that the network-reconstructed spatial structure and the hot plasma loop share a broadly consistent spatial distribution. It should be noted that AIA 1600~\AA\ traces transition-region and chromospheric UV radiation, which, while often spatially close to the HXI nonthermal bremsstrahlung, has a different physical origin; the use of AIA as a morphological reference in this section verifies the overall distributional consistency between the reconstructed bright region and the thermal/EUV loop structure, rather than providing an independent confirmation of the nonthermal electron deposition sites. Comparison with the AIA 171~\AA\ and 193~\AA\ bands further shows that the reconstructed hard X-ray bright region also exhibits a degree of spatial correspondence with the cooler loop structure and coronal loop emission, suggesting that the thermal and nonthermal radiation sources of this flare are broadly co-spatial.

\textbf{Limitations.} The validation in this section is based on a single flare event and is not yet sufficient to comprehensively evaluate the reconstruction performance of HXI-DLA2 across different flare classes and observing conditions. Background estimation uncertainty is the primary error source for this event: the background standard deviation is comparable to its mean, and subtraction errors directly affect the net-counts average energy and distribution. Furthermore, the training data are drawn from Hinode/XRT soft X-ray observations (thermal radiation), whereas the validation target here is HXI hard X-ray observations (nonthermal bremsstrahlung), which differ in radiation mechanism and source morphology; the reconstruction effectiveness of the network on this event suggests that the counts-to-spatial-distribution mapping it learns possesses a certain degree of cross-domain transferability, but this conclusion requires systematic verification on more events. The evolution of the flare source during the 16~s integration may also cause the net-counts distribution to deviate from the static source distribution assumption learned during training.

\section{Discussion and Summary}

\subsection{Discussion}

\textbf{Physical nature of the reconstruction.} A common feature across all the experiments above---the Gaussian morphologies, the double-source dynamic-range cases, and the soft X-ray complex sources---is that exact counts agreement does not imply pixel-level uniqueness---the compact-core example in the third row of Figure~\ref{fig:4_2_morphologies} illustrates this directly, with closely consistent counts despite a visibly deviating peak: the 91-dimensional counts compress the $161\times161$-pixel source into a far lower-dimensional measurement, making the inversion inherently underdetermined. The network output should therefore be understood as the \textbf{optimal prior-regularized solution consistent with the observed counts under the instrumental sampling constraint}, rather than a pixel-level replica of the true source. The forward equation confines the solution to the feasible range; the deep network selects the most reasonable spatial distribution within that range based on flare distribution regularities learned from training data.

\textbf{Data characteristics across the experiments.} Viewed together, the metrics in Tables~\ref{tab:gauss_compare}, \ref{tab:two_source_results}, \ref{tab:sxr_complex_results}, and \ref{tab:extreme_ratio_results} show a consistent pattern. On the compact Gaussian sources, the image-domain metrics are high (reconstruction--label correlation $R\gtrsim0.999$) and the re-projected counts agree closely with the input. On the complex soft X-ray morphologies, the image-domain metrics degrade on the training examples alike ($R$ down to $\sim$0.90 for the compact-core case; Table~\ref{tab:sxr_complex_results} training rows and Figure~\ref{fig:4_2_morphologies}) as on the temporally independent test events (Table~\ref{tab:sxr_complex_results}). The degradation therefore follows morphological complexity (loop-like and diffuse structures being the most difficult) rather than the train--test split, indicating that the limiting factor is the information content of the counts rather than overfitting to the training data. Within each case, the residuals concentrate where the counts are least informative: the fine structure of extended sources, the peak of the bright extended test event, and the faint component of the extreme-ratio double sources analyzed below.

\begin{figure*}[ht!]
\centering
\includegraphics[width=0.95\textwidth]{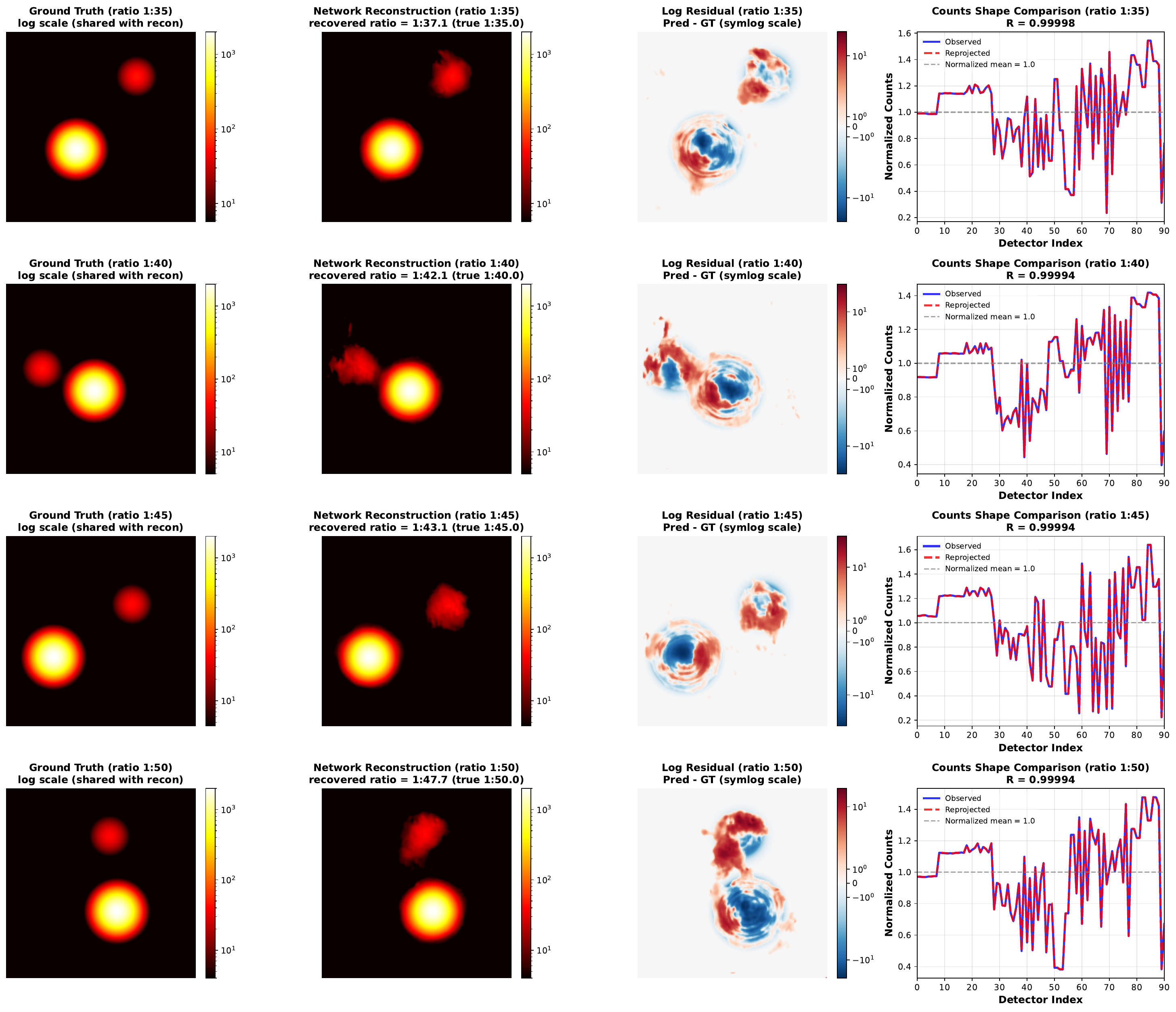}
\caption{Extreme dynamic-range double-source reconstruction results at peak ratios of 1:35, 1:40, 1:45, and 1:50. Each row corresponds to one peak intensity ratio. From left to right: the ground-truth label image, the network reconstruction (annotated with the recovered peak ratio), the logarithmic residual map (prediction minus label), and the counts-profile comparison between the observed and re-projected normalized counts (annotated with the correlation coefficient).}
\label{fig:extreme_peak_ratio}
\end{figure*}

\textbf{Extreme dynamic-range limit reconstruction.} To further probe the limits of the network's double-source resolution capability, we extend the dynamic-range experiments to the extreme regime of peak ratios from 1:35 to 1:50---well beyond the $\sim$20:1 level at which HXI\_CLEAN already loses extended weak sources \citep{Yu2025} and far above the $\sim$5:1 reached by the purely data-driven HXI-DLA. Figure~\ref{fig:extreme_peak_ratio} presents the reconstructions and Table~\ref{tab:extreme_ratio_results} summarizes the quantitative metrics: at every ratio the network successfully detects both sources, with NRMSE of order $10^{-3}$, SSIM above 0.995, and total-flux relative errors below 0.15\%. The flux ratios recovered via the peak-ratio method agree with the true ratios to within $\sim$7\% across the entire extreme range (e.g., 1:47.7 recovered for a true 1:50), even though the weak source contributes only $\sim$3\% of the total counts at 1:35 and $\sim$2\% at 1:50---a signature preserved in the normalized counts distribution and enforced through the distribution-consistency loss, so that double-source discrimination is anchored in the measurement domain rather than inferred from image priors alone. The residual image-domain errors are themselves informative about the ill-posedness of the inversion. The strong source is recovered almost exactly (peak error $\lesssim$2\%, centroid error $\lesssim$0.2~px), whereas the weak-source centroid error grows from $\sim$0.9~px at 1:35 to $\sim$2~px at 1:50 and its peak error reaches several percent, even changing sign at 1:45--1:50 where the weak peak is slightly overestimated---indicating residual fluctuations rather than a systematic bias. Embedded in the sidelobe field of the strong source, the weak source is the component most easily displaced by modulation fringes, and its $\sim$2\% counts contribution approaches the noise floor. These residual errors occupy precisely the pixel-space degrees of freedom that the 91-dimensional counts do not uniquely constrain---the re-projected counts remain consistent with the input, with counts-profile correlations above 0.9999 (Figure~\ref{fig:extreme_peak_ratio}), even though the image is not a pixel-perfect replica---illustrating that measurement-domain consistency and pixel-level fidelity are governed by different parts of the solution space. The physical constraints thus maintain effective double-source discrimination down to a weak-source signal of only $\sim$2\% of the total counts.

\begin{table*}[ht!]
\centering
\caption{Quantitative results of the extreme dynamic-range double-source experiments (peak ratios 1:35--1:50); metrics are as defined in Table~\ref{tab:two_source_results}.}
\label{tab:extreme_ratio_results}
\footnotesize
\setlength{\tabcolsep}{5pt}
\begin{tabular}{ccccccccc}
\toprule
Ratio & NRMSE & SSIM & Flux RE & Strong Peak RE & Weak Peak RE & Strong Cent & Weak Cent & $R$ \\
\midrule
1:35 & 0.0012 & 0.9980 & +0.08\%  & $-$0.34\% & $-$6.06\% & 0.06\,px & 0.94\,px & 0.99998 \\
1:40 & 0.0017 & 0.9966 & +0.11\%  & $-$1.22\% & $-$6.04\% & 0.12\,px & 1.77\,px & 0.99994 \\
1:45 & 0.0020 & 0.9969 & $-$0.05\% & $-$1.95\% & +2.49\%   & 0.15\,px & 1.96\,px & 0.99994 \\
1:50 & 0.0014 & 0.9954 & +0.09\%  & $-$0.74\% & +4.14\%   & 0.04\,px & 1.94\,px & 0.99994 \\
\bottomrule
\end{tabular}
\end{table*}

\textbf{Counts error analysis.} The counts errors of the network on real observations (RMSE $\sim$8.55\%) are markedly larger than on simulations (RMSE $<1$\%), arising mainly from the joint action of two factors. First, real HXI counts contain detector noise, cosmic-ray contamination, and residuals from background subtraction, whose inaccuracies are carried directly into the net counts feeding the network, whereas the training counts are noise-free mathematical projections. Second---and dominant---the non-negativity constraint $I(x,y)\ge 0$ enforced by HXI-DLA2 is physically correct, but net counts contaminated by noise and subtraction residuals may not be reproducible by any non-negative image: a perfect counts fit would require small negative values at background pixels, which the ReLU activation prohibits. Acting together with the input errors, this structural mismatch leaves an unfittable residual that dominates the elevated RMSE. In the training set, counts are computed from non-negative source images and are therefore always reproducible by a non-negative reconstruction; real observations do not share this property. Despite these factors, energy conservation and overall morphological recovery remain good, confirming that the physical prior constraint remains effective under realistic noise.

\textbf{Physical prior constraint versus purely data-driven learning.} In the earlier HXI-DLA study, a purely data-driven network performed reliably on compact sources but degraded on extended and complex morphologies, where the counts-to-image non-uniqueness is most severe. The present results corroborate that picture from the opposite direction: by confining the solution to the forward-equation-feasible subspace, HXI-DLA2 maintains stable quality precisely on the complex morphologies---ring sources, loop-like and diffuse morphologies, and extreme dynamic-range double sources---that are most challenging for unconstrained learning. The physical constraint reduces the effective search space of the inversion, so the deep network only needs to select the optimal solution within that subspace rather than learning the full unconstrained mapping. This confirms that combining physical constraints with data-driven priors is key to robust underdetermined inversion, a conclusion with general implications for modulation imaging.

\textbf{Methodological advantages of deterministic output.} HXI-DLA2 always produces a unique reconstruction given the same input, independent of manual parameter choices, eliminating subjective bias from parameter tuning. A single forward pass completes the reconstruction in milliseconds, suitable for real-time monitoring and batch processing.

\textbf{Limitations and applicability boundary.} The main limitations of the current model are: fixed patterns make the model region-specific, and full-disk coverage is a subject for future work; the real-data validation is based on a single flare event, and broader validation remains to be extended; the net-counts input inherits background-subtraction uncertainty.
\subsection{Summary and Outlook}

For the ASO-S/HXI hard X-ray imaging inversion problem, this paper builds upon our group's earlier HXI-DLA work and continues the development of image reconstruction methods for the HXI instrument. It establishes a counts average-energy--distribution decoupling relationship and, based on it, constructs a physics-constrained inversion network, HXI-DLA2. Taking the dirty image and the 91-dimensional counts as inputs, the network maps the decoupled average energy and distribution respectively to an energy-closure constraint at the output and a distribution-consistency constraint in the loss function, enforcing counts-average-energy closure at the output to guarantee energy conservation, while approximating the full-channel forward response through a distribution-consistency loss. The experiments show that HXI-DLA2 achieves reliable reconstruction on double-source and complex-morphology sources: double peaks are detected across dynamic ranges from 1:10 to the extreme 1:50---far beyond the $\sim$5:1 reached in our previous study---with counts-distribution correlation coefficients above 0.999; on temporally independent soft X-ray test events, the network reconstructs complex structures with NRMSE below 0.02, with the counts average energy matching the input exactly; on the real HXI observation, the reconstruction satisfies strict energy closure with the counts average energy matching the observation, and the reconstructed spatial morphology is consistent with contemporaneous SDO/AIA observations.

The core arguments of this work are fourfold. \textbf{First}, the counts average-energy--distribution decoupling (DC--AC decomposition) holds and can directly guide inversion network design. \textbf{Second}, reducing the solution space of the underdetermined inversion with the physical prior of the optical system while leveraging a deep network to learn the statistical regularities of flare source spatial distributions from training data and select the optimal solution within the solution space is an effective approach to underdetermined inversion; neither can be omitted. \textbf{Third}, the network reconstruction is the optimal prior-regularized solution under the instrumental sampling constraint, rather than a pixel-level replica of the true source. \textbf{Fourth}, we suggest that this inversion framework may be adaptable to other modulation-imaging instruments: embedding the optical sampling forward equation as a hard constraint while learning source-distribution regularities from training data is, in our view, a plausible strategy for any instrument that compresses spatial information into a finite set of modulation measurements, such as Solar Orbiter/STIX and RHESSI.

The long-standing difficulty that indirect modulation imaging collects only a limited number of spatial-frequency components, thereby constraining the ability to reconstruct complex sources \citep{Hurford2002, Krucker2020}, is partially alleviated by this work: the combined strategy of reducing the solution space with physical constraints and selecting the optimal solution within it via deep priors offers a viable path for recovering complex-source morphologies from finite modulation measurements.

Future work will embed the patterns as learnable parameters or conditional encodings to enable full-disk region transfer, and extend to pattern adaptation across different energy bands and systematic validation on more flare events.

%% Use the acknowledgment and contribution environments.
%% These will be anonymized when the "anonymous" style option is used.
\begin{acknowledgments}
This work is supported by the Natural Science Foundation of China (Nos. 12503097, 12373115), the Yunnan Science Foundation of China (No. 202501AT070025), the Yunnan Key Laboratory of Solar Physics and Space Science (No. 202205AG070009), and the Yunnan Revitalization Talent Support Program (Nos. 202305AS350029 and 202305AT350005). We thank the Astronomical Technology Laboratory, Yunnan Observatories for providing GPU computing resources. We thank the ASO-S/HXI, SDO/AIA, and Hinode/XRT teams for providing observational data. ASO-S (Kuafu-1) is a dedicated solar observation mission of the Chinese Academy of Sciences. The HXI-DLA2 code, trained models, and data generation pipeline are publicly available on ScienceDB (\url{https://doi.org/10.57760/sciencedb.46026}) under the Apache-2.0 license.
\end{acknowledgments}

\begin{contribution}
S.Z. and H.L. conceived the research idea and designed the experiments. S.Z. performed the theoretical analysis, network design, training, and experimental results. Y.S. provided the HXI modulation patterns and observational event data. J.H. provided guidance on the flare observation results. B.W. provided guidance on the principles of optical grid modulation. W.C. contributed to the refinement of the experimental scheme. K.J. and Z.J. provided guidance on the experimental design and multiple rounds of experimental validation. S.Z. wrote the manuscript. All authors reviewed and provided comments on the manuscript.
\end{contribution}

%% Following the acknowledgments section, use the \facilities{} macro to list
%% the keywords of facilities used in the research for the paper.
\facilities{ASO-S(HXI), SDO(AIA), Hinode(XRT)}

%% authors a place to specify which programs were used during the creation
%% of the manuscript.
\software{HXI-DLA2 (\url{https://doi.org/10.57760/sciencedb.46026})}

%% Bibliography generated with BibTeX plus aasjournalv7.bst
%% Compile sequence: pdflatex -> bibtex -> pdflatex -> pdflatex
\bibliography{references}{}
\bibliographystyle{aasjournalv7}

\end{document}